\documentclass[article,1p,11pt]{elsarticle}
\usepackage{lineno,hyperref}
\modulolinenumbers[5]
\usepackage[...]{youngtab}
\usepackage{xfrac}
\usepackage{appendix}
\usepackage{amsfonts}
\usepackage{amssymb}
\usepackage{physics}
\usepackage[usenames,dvipsnames]{xcolor}
\usepackage{tabularx,multirow,array,amsmath,mathalfa}
\usepackage{color}
\usepackage{ccaption}
\usepackage{mathtools}
\usepackage{bbold}
\usepackage{kantlipsum}
\usepackage{tikz}
\usepackage{pgfplots}
\newcommand{\mdot}{\raise1.5pt \hbox{.}}
\usepackage{braids}
\newtheorem{theorem}{Theorem}[section]
\newtheorem{theorem*}{Theorem}

\newtheorem{corollary}[theorem]{Corollary}

\newtheorem{defn}{Definition}[section]

\allowdisplaybreaks

\usepackage{changepage}
\usepackage[framemethod=TikZ]{mdframed}
\usepackage{lipsum}
\usepackage{titlesec}

\definecolor{Red}{rgb}{1,0,0}

\usepackage{pgf}
\usepackage{tikz}
\usepackage[utf8]{inputenc}
\DeclarePairedDelimiter\floor{\lfloor}{\rfloor}
\usetikzlibrary{arrows,automata}
\usetikzlibrary{positioning}
\tikzset{state/.style={rectangle, rounded corners, draw=black, very thick, minimum height=2em, inner sep=2pt, text centered,},}

\newcommand{\iu}{{i\mkern1mu}}

\journal{Nuclear Physics B}
\begin{document}

\begin{frontmatter}

\title{Colored HOMFLY-PT polynomials of quasi-alternating $3$-braid knots}
\author{Vivek Kumar Singh}
\ead{vivek\_s@uaeu.ac.ae}
\author{Nafaa Chbili}
\ead{nafaachbili@uaeu.ac.ae}
\address{Department of Mathematical Sciences, United Arab Emirates University,\\ Al Ain 15551 Abu Dhabi, UAE}

\begin{abstract}
Obtaining a closed form expression for the  colored HOMFLY-PT polynomials of  knots from $3$-strand braids carrying arbitrary $SU(N)$ representation is  a challenging problem. In this paper, we confine our interest to twisted generalized hybrid  weaving knots which we denote hereafter  by $\hat{Q}_3(m_1,-m_2,n,\ell)$. This family of knots not only  generalizes the well-known class of weaving knots but also    contains an infinite family   of quasi-alternating knots. Interestingly, we obtain  a closed form expression for the HOMFLY-PT polynomial of $\hat{Q}_3(m_1,-m_2,n,\ell)$ using a modified version of the Reshitikhin-Turaev method. In addition, we compute the exact coefficients of the Jones polynomials and the Alexander polynomials of quasi-alternating knots $\hat{Q}_3(1,-1,n,\pm 1)$. For these homologically-thin knots, such  coefficients are known to be  the ranks of their Khovanov and link  Floer homologies, respectively.  We also show  that the asymptotic behaviour  of the coefficients of the Alexander polynomial is  trapezoidal. On the other hand, we compute the  $[r]$-colored HOMFLY-PT polynomials of quasi alternating knots for small values of $r$. Remarkably, the study of the determinants of certain twisted weaving knots leads to establish a connection with enumerative geometry related to $m^{th}$ Lucas numbers, denoted hereafter as $L_{m,2n}$. At the end, we verify that the reformulated invariants  satisfy Ooguri-Vafa conjecture and we  express certain BPS integers in terms of hyper-geometric functions $\setlength\arraycolsep{1pt} {}_2 { \bf F}_1\left[\begin{matrix}a&, &b&, &c\end{matrix};z\right]$.

\end{abstract}

\end{frontmatter}

\newpage

\tableofcontents

\section{Introduction}
\begin{figure}[h]
\centering
\includegraphics[scale=.55]{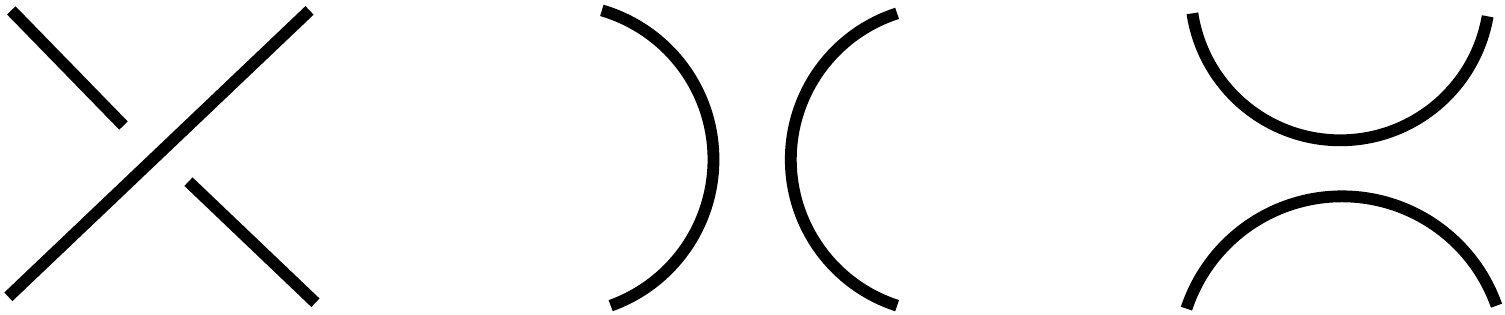}
$L$ \hspace{4.5cm} ~~~~~ ~$L_{0}$ \hspace{5.5cm}$L_{\infty}$
\caption{The diagram of the Link $L$ at the crossing $c$ and its smoothings $L_{0}$ and $L_{\infty}$, respecyively.}
\label{KBSM}
\end{figure}

Knot theory is the branch of low dimensional topology concerned  with the study of the spatial configurations of embedded circles in the 3-dimensional space.  These embeddings are considered up to natural deformations called isotopies. It is well known that the study of knots and links up to isotopies is equivalent to the study of their regular  planar projections up to some local moves. A link  projection is called alternating if the overpass and the underpass alternate as one travels along any component of the link. The class of alternating links is of central importance in classical knot theory and the study of the  Jones polynomials of these links  led to the solution of long-standing conjectures in knot theory \cite{Th}. Indeed, certain quantum invariants  reflect the alternating property of the link in a very obvious way. It was also proved that these links are homologically thin in both Khovanov and link Floer homology. Moreover, the Heegaard Floer homology of the branched double cover of an alternating link $L$ depends only on  the determinant of the link, ${\rm{Det}}(L)$. These interesting homological properties extend to a wider class of links called quasi-alternating links. Unlike alternating links which admit a simple diagrammatic description, quasi-alternating  links are defined in the following  recursive way.
\begin{defn}\label{defQA}
The set $\mathcal{Q}$ of quasi-alternating links is the smallest set
satisfying the following properties:
\begin{enumerate}
    \item The unknot belongs to $\mathcal{Q}$.
  \item If $L$ is a link with a diagram $D$ containing a crossing $c$ such that
\begin{enumerate}
\item both smoothings  of the diagram $D$ at the crossing $c$, $L_{0}$ and $L_{\infty}$
as in Figure \ref{KBSM} belong to $\mathcal{Q}$;
\item $\det(L_{0}), \det(L_{\infty}) \geq 1$;
\item $\det(L) = \det(L_{0}) + \det(L_{\infty})$; \\
then $L$ is in $\mathcal{Q}$.
In this case we say that  $L$ is quasi-alternating  with quasi-alternating diagram $D$ at the
crossing $c$.
\end{enumerate}
\end{enumerate}
\end{defn}
With an elementary induction on the determinant of the link, this   definition   can be used to prove that non-split alternating links are quasi-alternating. The knots $8_{20}$ and $8_{21}$ are the first examples, in the knot table,  of  non-alternating quasi-alternating knots.  In general, using the definition above to decide   whether a given link is quasi-alternating is a very  challenging task.\\
 
For $m>1$, let $\mathcal{B}_m $ be the group of braids on $m$ strands. It is well known that any link in the 3-dimensional space can be represented as the closure of an $m$ braid. The smallest such integer $m$ is called the braid index of the link. Links of braid index $2$ are torus links.  Links of braid index $3$ have been subject to extensive study. Indeed, braids on 3 strands  have been classified, up to conjugacy, into normal forms, see \cite{Mu}. This classification led to a better understanding of 3-braid links. In particular, alternating links of braid index 3 have been classified by Stoimenow \cite{St}.  Quasi-alternating links of braid index $3$ have been characterized  by Baldwin \cite{Ba}. Other than a finite number of links, any quasi-alternating link of braid index three is the closure of a braid of the form
$\sigma_1^{p_1}\sigma_2^{-q_1}\dots \sigma_1^{p_n}\sigma_2^{-q_n}h^{\ell}$ where $p_i,q_i$ are positive integers,  $h$ is the central braid $(\sigma_1\sigma_2)^3$, and  $\ell=0,1,$ or $-1$. If $p_i=m_1$ for all $ i$ and $q_i=m_2$ for all $i$ then this link is denoted by $\hat{Q}_3(m_1,-m_2,n,\ell)$. These links are refereed to as twisted generalized hybrid  weaving knots.   Note that the case $m_1=m_2=1$ and $\ell=0$, corresponds to  the classical  class of weaving links. If $m_1=m_2$ and $\ell=0$, then we obtain the hybrid weaving links whose colored HOMFLY-PT polynomials  have been recently studied in \cite{VRR}. One may easily see that  these hybrid weaving links are alternating.
The purpose of this paper is to extend the work in \cite{VRR} to a wider class of links  which in particular includes a large family of   links which are quasi-alternating  but not alternating.  More precisely, we  study the colored HOMFLY-PT polynomial of the links  $\hat{Q}_3(m_1,-m_2,n,\ell) $ using a modified version of the Reshitikhin-Turaev method. Our main result is an explicit formula for the HOMFLY-PT polynomials of these class of links.\\

Here is an outline of the paper.  In Section \ref{s.int}, we shall review the modified version of Reshitikhin-Turaev  method for constructing knot invariants and describe  the block structure form of  $\mathfrak {R}$-matrices in the case of   three-strand braids. In Section \ref{s3.int}, we use  the twisted trace $\Psi^{[2,1,0]}((m_1,-m_2,n,\ell);A,q)$, to  introduce  a closed form expression for  the   HOMFLY-PT polynomial of the knot $\hat{Q}_3(m_1,m_2,n,\ell) $. As a consequence, we show that  the determinants of hybrid weaving knots are related to Lucas numbers. Further, we prove the trapezoidal behaviour of the coefficients of the Alexander polynomial of weaving knots.  Section \ref{s4.int} is devoted to study  the  $[r]$-colored HOMFLY-PT polynomial of  quasi-alternating knots $\hat{Q}_3(1,-1,n,\pm 1)$. In particular,  the twisted traces of 2-dimensional matrices are expressed   as  Laurent polynomials $\Theta_{n,t,m,\ell}[q]$ and the  colored HOMFLY-PT polynomials of  quasi-alternating knots are calculated. In Section \ref{s5.int}, we verify that the reformulated invariants from these quasi-alternating knot invariants satisfy Ooguri-Vafa Conjecture.  Section \ref{s6.int} summarizes the paper and  discusses  related challenging open problems.
Finally,  explicit data on colored HOMFLY-PT polynomials up to representation $[r]=[4]$ as well as reformulated invariants for $\hat{Q}_3(1,-1,n,\pm 1)$ are given in Appendices \ref{app0}, \ref{app1} and \ref{app2}.

\section{Knot invariants from quantum groups}\label{s.int}

A fundamental  result in classical knot theory states that  any link in the 3-dimensional sphere can be viewed as the closure of a braid on $m$ strands. This fact  has played a key role in the development of knot theory. In particular, quantum knot invariants are  constructed from representations of the braid group. Such a representation
associates a  quantum $\mathfrak{R}$-matrix to each of the generators of the braid group  $\sigma_1,\ldots,\sigma_{m-1}$, see \cite{KirResh} and \cite{Rosso:1993vn, lin2010hecke, Liu:2007kv}:

\begin{equation}
\begin{array}{rcl}
\pi: \mathcal{B}_m  &\rightarrow&  \text{End}(V_1\otimes\ldots,\otimes V_m) \\
\pi(\sigma_{\mu}) &=& \mathfrak{R}_{\mu}.
\end{array}
\end{equation}


These matrices has to satisfy  the following relations:
\begin{eqnarray}\label{conjugation}
{\mathfrak {R}}_{\mu}{{\mathfrak {R}}}_{\nu}&=&{{\mathfrak {R}}}_{\nu}{{\mathfrak {R}}}_{\mu}~~ \text{for}~ |{\mu}- {\nu}|>1~,
\end{eqnarray}
\begin{eqnarray}\label{braidp}
{{\mathfrak {R}}}_{\mu}{{\mathfrak {R}}}_{{\mu}+1}{{\mathfrak {R}}}_{{\mu}} &=&{ {\mathfrak {R}}}_{{\mu}+1}{{\mathfrak {R}}}_{\mu}{{\mathfrak {R}}}_
{{\mu}+1},~\text{for}~\ {\mu}=1,\ldots,m-2.
\end{eqnarray}

Note that the matrix $\mathfrak {R}_{\mu}$ acts on the  two consecutive braid stands $\mu$ and $\mu+1$.
Pictorially,  $\mathfrak {R}_{\mu}$  is  represented  as follows:
\begin{equation}\label{RM}
\begin{picture}(850,140)(-250,-90)

\put(-70,20){\line(0,-1){90}}
\put(-40,20){\line(0,-1){90}}

\put(0,20){\color{blue} \line(0,-1){30}}
\put(30,20){\color{blue}\line(0,-1){30}}
\put(0,-40){\color{blue}\line(0,-1){30}}
\put(30,-40){\color{blue}\line(0,-1){30}}
\put(70,20){\line(0,-1){90}}

\put(0,-10){\color{blue}\line(1,-1){30}}
\multiput(30,-10)(-17,-17){2}
{\color{blue}\line(-1,-1){13}}

\put(-105,-25){\mbox{$\mathfrak {R}_{\mu}$ \ = \ }}
\put(45,-25){\mbox{$\ldots$}}
\put(-25,-25){\mbox{$\ldots$}}
\put(-75,25){\mbox{$ V_1$}}
\put(-45,25){\mbox{$V_2$}}
\put(-5,25){\mbox{$\color{blue}V_{\mu}$}}
\put(20,25){\mbox{$\color{blue}V_{\mu+1}$}}
\put(60,25){\mbox{$ V_m$}}
\end{picture}
\end{equation}
In 1984, Jones introduced a topological invariant of oriented links through the study of representations of the braid group via  certain  von Neumann algebras. The Jones polynomial was later generalized to a two-variable invariant  known as the HOMFLY-PT polynomial. Following the seminal work of Witten  \cite{Wit}, colored versions of these invariants have been defined   by Reshetikhin and Turaev   and used to construct  new quantum invariants of 3-manifolds \cite{RT1, RT2}.  In this paper, we are interested in the so-called colored HOMFLY-PT polynomial $H_{[
r]}^{\mathcal{K}} (q, A)$; a  two-variable polynomial  defined as follows
$$H_{[
r]}^{\mathcal{K}} (q, A)=  {}
_q\text{tr}_{V\otimes\dots\otimes V}\left( \, \pi(\beta_{\mathcal{K}}) \, \right),$$
 where ${}_q\text{tr}_V(z)$ denotes  the quantum trace  of $z$ \cite{ Klimyk} and $\beta_{\mathcal{K}}$ is a braid word whose closure is the  knot $\mathcal{K}$. The  Reshetikhin-Turaev approach, denoted hereafter  RT for short,   involves  non-diagonal universal $\check R$ matrices  that makes the computations of knot invariants  somehow  cumbersome. A more effective method is discussed in \cite{ModernRT1,Mironov:2011ym, Anokhina:2013wka}, where the colored HOMFLY-PT polynomials are viewed  in the form of character decomposition. This formalism is  known as the  modified RT-approach. The key feature in this approach is that the braiding generators can be expressed in a block structure form. This allows a better control of the computation of knot invariants. In this article, we use modified RT-approach to compute the  symmetric $[r]$-colored HOMFLY-PT polynomials for {\it twisted} generalized hybrid weaving knots.

\subsection{ The modified Reshetikhin-Turaev approach}\label{s11.int}
\begin{figure}[h]
\centering
\includegraphics[scale=.6]{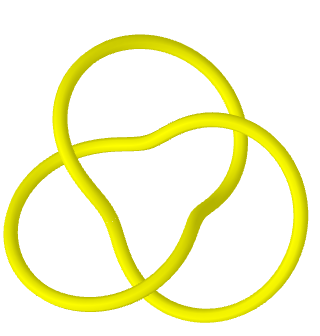}
\caption{ Trefoil knot ${\bf 3_{1}}$.}
\label{TK}
\end{figure}

In this section, we shall briefly summarize the modified RT-approach. More technical details  can be found in  \cite{Mironov:2011ym, ModernRT1,Anokhina:2013wka, Dhara:2018wqe,VRR}.  It is worth mentioning here   that we are assuming that  each strand of the braid is  labelled  with the same   representation $({\bf R})$\footnote{Note that all finite dimensional representations of $U_q(sl_N)$ can be enumerated using Young diagrams. The conventional notation  $[\alpha_1,\alpha_2,\alpha_3,\ldots, \alpha_{N-1}]$ to denote the young diagram. Here,  $\alpha_1$ boxes in the first row, $\alpha_2$  boxes in the second row and so on. For instance, $[4,2,0,\ldots]={\tiny\yng(4,2)}$.}. This method involves $\hat{\mathfrak{R}}_{\mu}$-matrices with block structure form $$\hat{\mathfrak{R}}=diag\{\hat{\mathfrak{R}}^{\Xi_{\alpha}}\},$$ which arises from the tensor product decomposition of  symmetric representations $({\bf R})$:
\begin{eqnarray}\label{irrdec}
 {{\bf R}^{\bigotimes}}^{m} &=& \bigoplus_{\alpha,~\Xi_{\alpha} \vdash m |{\bf R}|} ({\rm dim} {\mathcal{M}}^{1,2 \ldots m}_{\Xi_{\alpha}}) ~ \Xi_{\alpha}~,
\end{eqnarray}
where $\Xi_{\alpha}$ denotes the irreducible representation labeled by index $\alpha$ and $\mathcal{M}_{\Xi_{\alpha}}^{1,2, \ldots, m}$ captures the  multiplicity of an irreducible representation. So, for an  $m$-strand braid, we have $m-1$ different $\mathfrak{R}_{\mu}$ matrices. Therefore, $\mathfrak{R}^{\Xi_{\alpha}}_{\mu}$ can be diagonalized but not simultaneously for all $\mu$. Moreover, these different matrices are related by conjugation:
\begin{equation}
\hat{\mathfrak{R}}_{\mu}^{\Xi_{\alpha}}=\mathcal{U}^{\Xi_{\alpha}}_{\mu \nu}\hat{\mathfrak{R}}_{\nu}^{\Xi_{\alpha}}(\mathcal{U}^{ \Xi_{\alpha}}_{\mu \nu})^{\dagger}.
\end{equation}
Here, the symbol $\dagger$ stands for  the conjugate-transpose of the matrix
and  $\mathcal{U}$ is a unitary matrix having same block diagonal form,   i.e.,
$$\mathcal{U}_{\mu \nu}=diag\{\mathcal{U}^{\Xi_{\alpha}}_{\mu \nu}\}.$$
In \cite{Mironov:2011ym}, the normalized colored HOMFLY-PT polynomials of a knot $\mathcal{K}$ is defined as follows:
\begin{eqnarray}
H_{\bf R}^{ \mathcal{K}} \left(q,\,A=q^N \right)
&=& \frac{1}{S^{*}_{\bf R}}\sum\limits_{\alpha,~\Xi_{\alpha} \vdash m |\bf R|} S^{*}_{\Xi_{\alpha}}(A,q){h}^{\Xi_{\alpha}}_{\bf R}(q),
\label{HOMFLY}
\end{eqnarray}
where $\Xi_{\alpha}$ represents the  irreducible representation in the product ${\bf R}^{\otimes m}$, $m$ stands for the number of braid strands, ${\bf R}$ denotes the representation on each strand and $S^*_{\Xi_{\alpha}}(A,q)$  are the Schur functions (characters of the linear groups $GL(N)$). The coefficient ${h}^{\Xi_{\alpha}}_{\bf R}(q) $ in (\ref{HOMFLY})  depends on the  variable $q$ and is defined as follows, see  \cite{morozov2010chern}:
 $${h}^{\Xi_{\alpha}}_{\bf R
}(q)={\rm Tr}_{\mathcal{M}^{1,2 \ldots m}_{\Xi_{\alpha}}}C_{\beta_{\mathcal{K}}}.$$
Here $C_{\beta_{\mathcal{K}}}$ can be expressed in  terms  of a product of $\hat{\mathfrak{R}}$-matrices which correspond to 2-dimensional  projection of  the knot $\mathcal{K}$. In order to define  a quantum trace of $C_{\beta_{\mathcal{K}}}$, one needs to  define a basis states in weight space incorporating the multiplicity as well, i.e.,
 \begin{equation}
\label{decomp}
\vert \left(\ldots \left(( {\bf R} \otimes {\bf R})_{\Lambda_{\alpha}} \otimes {\bf R}\right)_{{\Xi}_{\alpha_1}} \ldots {\bf R}\right)_{\Xi_{\alpha}}\rangle^{(\mu)} \equiv | \Xi_{\alpha}; \Xi_{\alpha,\mu}, \Lambda_{\alpha}\rangle
\cong~| \Xi_{\alpha,\mu},\Lambda_{\alpha}
\rangle \otimes |\Xi_{\alpha}\rangle~,
\end{equation}
where $\Lambda_{\alpha}\in {\bf R}^{\otimes 2}$ and $\mu$ keeps track of the multiplicity. It is a good choice of eigen state of quantum
$\hat{\mathcal{ R}}^{\Xi_{\alpha}}_1$ matrix that satisfies
\begin{eqnarray}
\hat{\mathfrak{R}}^{\Xi_{\alpha}}_1 \vert \Xi_{\alpha}; \Xi_{\alpha,\mu}, \Lambda_{\alpha} \rangle =
\lambda_{\Lambda_{\alpha},\mu}({\bf R},{\bf R})\vert \Xi_{\alpha}; \Xi_{\alpha,\mu}, \Lambda_{\alpha} \rangle \nonumber.
\end{eqnarray}
It is clear that  the matrix $\hat{\mathfrak R}^{\Xi_{\alpha}}_1$ is diagonal in the basis $\vert \Xi_{\alpha}; \Xi_{\alpha,\mu}, \Lambda_{\alpha} \rangle $. According to \cite{Klimyk,GZ}, for a symmetric representation ${\bf R}=[r]$, the explicit  values of the braiding eigenvalues  are given by:
 \begin{equation}
 \label{evR1}
       \lambda_{\Lambda_{\alpha},\mu} ([r],[r])=
            \epsilon_{\Lambda_{\alpha},\mu} A^{-r}q^{-f_{r}}q^{\varkappa(\Lambda_{\alpha})-
4\varkappa([r])}~,
        \end{equation}
where $f_{r}=2r(r-1)$ and $\varkappa(\Lambda_{\alpha})=\tfrac{1}{2} \sum_j \alpha_j (\alpha_j+1-2j)$ is cut-and-join-operator eigenvalue \cite{MMN,MMN1} of Young tableaux representation $\Lambda_{\alpha}=[2r-\alpha,\alpha]$ and $\epsilon_{\Lambda_{\alpha},\mu}$ is $\pm 1$.\footnote {The multiplicity subspace state $\Xi_{\alpha,\mu}$  is connected by $\Lambda_{\alpha}$ and zero otherwise.} The other $\hat{\mathfrak{R}}^{\Xi_{\alpha}}_{\mu}$-matrices can be determined by Equation \ref{conjugation}.
Now, we shall illustrate this construction by computing  the  $[r]$-colored HOMFLY-PT  polynomial of  the  trefoil knot ${\bf 3_{1}}$(see Figure \ref{TK}). Let us consider the trefoil knot as the closure  of the  2-braid $\sigma_1^{3}$, (\ref{2st}),  carrying symmetric representation $[r]$.  For this case, $\Xi_{\alpha}=\Lambda_{\alpha} \in [r]\bigotimes[r]$ and it has no multiplicity. Hence, the $\hat{\mathcal R}$ matrices are one-dimensional.

\begin{equation}
\hskip0cm
\begin{tikzpicture}{\label{2st}}
\braid[number of strands=2,rotate=90,green] (braid) a_1 a_1 a_1;
\end{tikzpicture}
\end{equation}

From Equation \ref{HOMFLY} and Equation \ref{evR1}, we have  \begin{eqnarray}{\label{egn}}
C_{\beta_{\bf 3_{1}}}&=&\hat{\mathfrak{R}}_1^3,\nonumber\\
\lambda_{\Lambda_{\alpha}}([r],[r]) &=& (-1)^{\alpha}A^{- r}q^{(\alpha^2-\alpha-2r \alpha+r)}.
\end{eqnarray}
 Hence the colored HOMFLY-PT polynomials  of  trefoil knot  are given by
\begin{eqnarray}\label{HOMR}
H_{[r]}^{\mathcal{K}}(q,A)&=&\frac{1}{ S^*_{[r]}}\sum_{\alpha}  {\rm Tr}_{\mathcal{M}^{1,2
}_{\Xi_{\alpha}}}(\hat{\mathfrak{R}}_1)^3~ S^*_{ \Lambda_{\alpha}}=\frac{1}{ S^*_{[r]}}\sum_{\alpha} \lambda_{\Lambda_{\alpha}}([r],[r])^{3} ~ S^*_{ \Lambda_{\alpha}},\nonumber\\
&=&\frac{A^{-3 r}}{ S^*_{[r]}}\sum_{\alpha=0}^{r} (-1)^{3\alpha} q^{3 (\alpha^2-\alpha-2r \alpha+r)}
~S^*_{ \Lambda_{\alpha}}.
\end{eqnarray}
 Here $S^*_{ \Lambda_{\alpha}}$ is the  quantum dimension of $\Lambda_{\alpha}$.\footnote{ The explicit form of  the quantum dimension is
 \begin{equation*}
S^*_{\Lambda_{\alpha}} = \frac{[N+\alpha-2]_q!\,[N+2r-\alpha-1]_q!\,[2r-2\alpha+1]_q}{[\alpha]_q!\,[2r-\alpha+1]_q!\,[N-1]_q!\,[N-2]_q!},~~~ [n]_{q} = \frac{q^{n} - q^{-n}}{q^{1} - q^{-1}},
\end{equation*}
where $[n]_{q}! = \prod_{i=1}^n [i]_q$ with $[0]_{q}!=1$.}
The explicit colored polynomials of  the trefoil knot ${\bf 3_1}$ for  $[r]=[1], [2]$ are given  below:
\begin{eqnarray*}
H^{{\bf 3_1}}_{[1]}(q,A)&=&\frac{A^2}{q^2}(1 - A^2q^2 + q^4),\\
H^{{\bf 3_1}}_{[2]}(q,A)&=&\frac{A^4}{q^4}(1 - A^2 q^4 + q^6 - A^2 q^6 + q^8 - A^2 q^{10} + A^4 q^{10} + q^{12 }-
  A^2 q^{12}).
\end{eqnarray*}
  Note that as we  move to  the study of braids with higher number of strands, multiplicity structures may arise in the quantum $\hat{\mathfrak{R}}_i$-matrices. We shall illustrate the steps of the modified RT-approach  for three-strand braids in the following section.

\subsection{ $\hat{\mathfrak{R}}$-matrices for three-strand braids}\label{s12.int}
\begin{figure}[h]
\centering
\includegraphics[scale=.5]{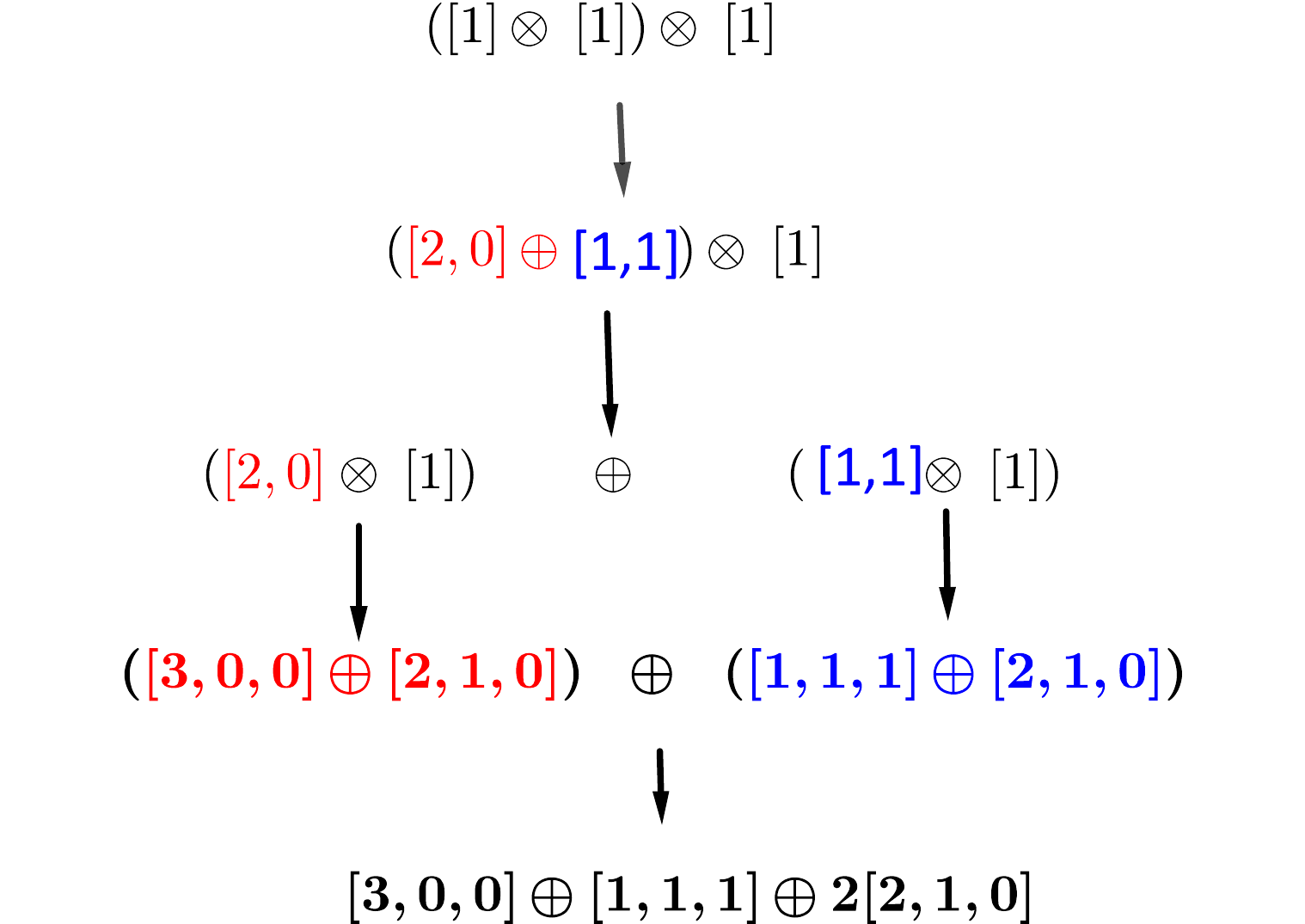}
\caption{ Path and block structures of irreducible representations $[3,0,0]$, $[1,1,1]$ and $[2,1,0]$.}
\label{TK1}
\end{figure}

We consider a braid with $3$- strands  each of which is  associated with a symmetric representation $[r]$. Then, the tensor product decomposition is given as follows:
$$\bigotimes^3 [r] =\bigoplus_{\alpha}({\rm dim} {\mathcal{M}}^{1,2,3}_{\Xi_{\alpha}})\Xi_{\alpha},$$
here the Young tableaux irreducible representations $\Xi_{\alpha}\equiv[{\xi_1}^{\alpha},{\xi_2}^{\alpha},{\xi_3}^{\alpha}]$ are such that ${\xi_1}^{\alpha}+{\xi_2}^{\alpha}+{\xi_3}^{\alpha}=3 r$ and ${\xi_1}^{\alpha}\geq {\xi_2}^{\alpha}\geq{\xi_3}^{\alpha} \geq 0 $. For example
\begin{eqnarray*}
\label{irrdec}
\bigotimes^3 [1] &=& [3,0,0]\bigoplus [1,1,1]\bigoplus 2 [2,1,0].
\end{eqnarray*}
First, we discuss the  path and block structure of $\hat{\mathfrak{R}}$-matrix for irreducible representations $[3,0,0]$, $[1,1,1]$ and $[2,1,0]$ shown in Figure \ref{TK1}. Note that the multiplicity of the representations $[3,0,0]$, $[1,1,1]$, and $[2,1,0]$ is equal to one, one, and two respectively.
Since the braid group $\mathcal{B}_3$ involves two braiding generators, i.e., $\hat{\mathfrak {R}}_1$ and $\hat{\mathfrak {R}}_2$-matrices. The $\hat{\mathfrak{R}}^{[2,1,0]}_1$-matrix is determined from the eigenvalue equation (\ref{evR1}):
 \begin{eqnarray}
 \label{R1}
       \hat{\mathfrak{R}}^{[2,1,0]}_1| [2,1,0]_{1},[2,0,0]
;[2,1,0]\rangle &=& \frac{q}{A}| [2,1,0]_{1},[2,0,0]
;[2,1,0]\rangle ,\\~         \hat{\mathfrak{R}}^{[2,1,0]}_1| [2,1,0]_{2},[1,1,0]
;[2,1,0]\rangle&=& -\frac{1}{A q}| [2,1,0]_{2},[1,1,0]
;[2,1,0]\rangle\rangle.
\end{eqnarray}
Hence, the explicit form of $\mathfrak{R}_1$ is
\begin{equation} \label{R12}
\mathfrak{R}^{[2,1,0]}_1= \frac{1}{A}\left(\begin{array}{cc} q & 0 \\ \\  0 & -\frac{1}{q} \end{array}\right),
\end{equation}
Similarly,
 \begin{eqnarray}
 \label{R1}
       \hat{\mathfrak{R}}^{[3,0,0]}_1| [3,0,0]_1,[2,0,0]
;[3,0,0]\rangle&=& \frac{q}{A}| [3,0,0]_1,[2,0,0]
;[3,0,0]\rangle,\\~         \hat{\mathfrak{R}}^{[1,1,1]}_1| [1,1,1]_1,[1,1,0]
;[1,1,1]\rangle&=& -\frac{1}{A q}| [1,1,1]_1,[1,1,0]
;[1,1,1]\rangle.
\end{eqnarray}
Note that $\hat{\mathfrak {R}}_1$ and $\hat{\mathfrak {R}}_2$ cannot be simultaneously diagonal but related by a unitary matrix which can be identified with the $U_q(sl_N)$ Racah matrix discussed in details in \cite{Itoyama:2012re, Dhara:2017ukv, Dhara:2018wqe}. Therefore, from Equation \ref{conjugation}, $\hat{\mathfrak{R}}_2$ is defined as
\begin{equation*}
\hat{\mathfrak{R}}_2^{\Xi_{\alpha}}=\mathcal{U}^{\Xi_{\alpha}}\hat{\mathfrak{R}}_1^{\Xi_{\alpha}}(\mathcal{U}^{ \Xi_{\alpha})^{\dagger}}.
\end{equation*}
Here $\mathcal{U}^{\Xi_{\alpha}\dagger}$ denotes the conjugate-transpose of $\mathcal{U}^{\Xi_{\alpha}}$. Algebraically the  matrix $\mathcal{U}^{\Xi_{\alpha}}$ relates two equivalent basis states as shown below:
\begin{equation*}
\label{asis}
|\left(\left( [r]\otimes [r]\right)_{\Lambda_{\alpha}}\otimes [r]\right)_{\Xi_{\alpha}} \rangle \xrightarrow[\text{}]{\mathcal{U}^{\Xi_{\alpha}}} \vert \left([r] \otimes \left( [r]\otimes [r] \right)_{\Lambda_{\alpha'}} \right)_{\Xi_{\alpha}}\rangle~.
\end{equation*}
 The complete map of $\mathcal{U}^{\Xi_{\alpha}\equiv [\xi_{1}^{\alpha},\xi_{2}^{\alpha},\xi_{3}^{\alpha}]}$ matrix to $U_q(sl_2)$ Racah matrix is  discussed in details in \cite{Dhara:2018wqe},  we have
\begin{eqnarray}
\label{symrac}
\mathcal{U}^{\Xi_{\alpha}\equiv [\xi_{1}^{\alpha},\xi_{2}^{\alpha},\xi_{3}^{\alpha}]}
&=&
U_{U_q(sl_2)}
\begin{bmatrix}
(r-{\xi_3}^{\alpha})/2 & (r-{\xi_3}^{\alpha})/2 \\
~&~\\
(r-{\xi_3}^{\alpha})/2 & ({\xi_1}^{\alpha}-{\xi_2}^{\alpha})/2
\end{bmatrix}.
\end{eqnarray}
The closed form expression of $U_q(sl_2)$ Racah coefficients can be found in  \cite{KirResh}.
From Equation  \ref{symrac}, the explicit form of unitary matrix $\mathcal{U}^{[2,1,0]}$ is
\begin{eqnarray}
\mathcal{U}^{[2,1,0]}&=&U_{U_q(sl_2)}
\begin{bmatrix}
\frac{1}{2} &\frac{1}{2} \\
~&~\\
\frac{1}{2} & \frac{1}{2}
\end{bmatrix},\nonumber\\
&=& \left(\begin{array}{cc} \frac{1}{[2]_{q}} & \frac{\sqrt{[3]_q}}{[2]_{q}}\\ \\ -\frac{\sqrt{[3]_q}}{[2]_{q}}& \frac{1}{[2]_{q}} \end{array}\right).~~
\end{eqnarray}
Hence, the $\hat{\mathfrak{R}}_{2}$ matrices are
\begin{eqnarray}{\label{R2}}
\hat{\mathfrak{R}}^{[2,1,0]}_2&=&\mathcal{U}^{[2,1,0]} \hat{\mathfrak{R}}^{[2,1,0]}_1\mathcal{U}^{[2,1,0]\dagger},\nonumber\\&=&\frac{1}{A}\left(\begin{array}{cc} \frac{q^2 - [3]_q}{q {[2]_q}^2} &- \frac{\sqrt{[3]_q}}{[2]_{q}}\\ \\ - \frac{\sqrt{[3]_q}}{[2]_{q}}&  -\frac{1-q^2 [3]_q}{q {[2]_q}^2}  \end{array}\right).
\end{eqnarray}
Similarly,
\begin{eqnarray}{\label{R21}}
\hat{\mathfrak{R}}^{[3,0,0]}_2&=& \hat{\mathfrak{R}}^{[3,0,0]}_1=\frac{q}{A},\nonumber\\
\hat{\mathfrak{R}}^{[1,1,1]}_2&=&\hat{\mathfrak{R}}^{[1,1,1]}_1=-\frac{1}{qA}.
\end{eqnarray}
With this description of the RT-approach for 3-stand braids, we will investigate the polynomial invariants of twisted generalized hybrid weaving knots $\hat{Q}_3(m_1,-m_2, n,\ell)$ in the next  section.

\section{Twisted generalized hybrid weaving knots $\hat{Q}_3(m_1,-m_2, n,\ell)$}\label{s3.int}
In this section, we shall study the HOMFLY-PT polynomial of  {\it twisted} generalized hybrid weaving knots. These knots,
denoted hereafter as $\hat{Q}_3(m_1,-m_2, n,\ell)$,  are  obtained as the closure of three-strand braids $$(\sigma_1^{m_1} \sigma_2^{-m_2})^n\underbrace{(\sigma_1 \sigma_2)^{ 3 \ell}}_{\color{red}twist},$$
where $m_1,m_2, n$ are positive integers and $\ell \in \mathbb{Z}$(see in picture \ref{GWK}). Notice, that without loss of generality we may assume that $m_1 \geq m_2$.\\
\begin{equation}
\begin{picture}(350,100)(40,-50)\label{GWK}

%
\put(75,-30){\vector(1,0){80}}
%
%
\put(75,30){\vector(1,0){40}}
\put(75,0){\vector(1,0){40}}
\qbezier(115,30)(120,30)(120,25)
\qbezier(115,0)(120,0)(120,5)
\put(115,25){\line(1,0){20}}
\put(115,5){\line(0,1){20}}
\put(120,12){\mbox{\small $m_1$ }}
\put(115,5){\line(1,0){20}}
\put(135,5){\line(0,1){20}}
\qbezier(130,25)(130,30)(135,30)
\qbezier(130,5)(130,0)(135,0)
\put(135,30){\vector(1,0){20}}
\put(135,0){\vector(1,0){20}}
\qbezier(155,-30)(160,-30)(160,-25)
\qbezier(155,0)(160,0)(160,-5)
\put(155,-25){\line(1,0){20}}
\put(155,-5){\line(0,-1){20}}
\put(157,-17){\mbox{\small -$m_2$}}
\put(155,-5){\line(1,0){20}}
\put(175,-5){\line(0,-1){20}}
\qbezier(170,-25)(170,-30)(175,-30)
\qbezier(170,-5)(170,0)(175,0)
\put(155,30){\line(1,0){40}}
\put(175,-30){\vector(1,0){60}}
\put(175,0){\vector(1,0){20}}
\qbezier(195,30)(200,30)(200,25)
\qbezier(195,0)(200,0)(200,5)
\put(195,25){\line(1,0){20}}
\put(195,5){\line(0,1){20}}
\put(200,12){\mbox{\small $m_1$}}
\put(195,5){\line(1,0){20}}
\put(215,5){\line(0,1){20}}
\qbezier(210,25)(210,30)(215,30)
\qbezier(210,5)(210,0)(215,0)
\put(215,30){\vector(1,0){60}}
\put(215,0){\vector(1,0){20}}
\qbezier(235,-30)(240,-30)(240,-25)
\qbezier(235,0)(240,0)(240,-5)
\put(235,-25){\line(1,0){20}}
\put(235,-5){\line(0,-1){20}}
\put(236,-17){\mbox{\small -$m_2$}}
\put(235,-5){\line(1,0){20}}
\put(255,-5){\line(0,-1){20}}
\qbezier(250,-25)(250,-30)(265,-30)
\qbezier(250,-5)(250,0)(255,0)
\put(255,-30){\vector(1,0){20}}
\put(255,0){\vector(1,0){20}}

\put(285,0){\mbox{$\ldots n$}}

\put(310,-29){\begin{tikzpicture}
\braid[number of strands=2,rotate=90] (braid) a_1 a_2 ;
\end{tikzpicture}}
\put(400,0){\mbox{$\ldots 3\ell$}}
\end{picture}
\end{equation}

It is noteworthy that the link $\hat{Q}_3(m_1,-m_2, n,\ell)$ is alternating  if $\ell=0$. It is quasi-alternating non-alternating if $\ell=\pm 1$.   A few examples of knots of type  $\hat{Q}_3(m_1,-m_2, n,\ell)$ are given in Table \ref{table1}  below.\\
\begin{table}[h]
\begin{center}
\begin{tabular} { | p {5 cm} | p {7 cm}| }
\hline
Notation & Knot \\
\hline
$\hat{Q}_3(3,-1,1,-1)$ & $5_{2}$ Knot \\
$\hat{Q}_3(5,-1,1,-1)$ & $8_{20}$ Knot \\
$\hat{Q}_3(1,-1,n,0)$ & \text{weaving knot} ${W}(3, n)$ \\
$\hat{Q}_3(m,-m,n,0)$ & \text{hybrid weaving knot} $\hat{W}_3(m,-m, n)$ \\
\hline
\end{tabular}
\caption{Examples of twisted generalized hybrid weaving knots $\hat{Q}_3(m_1,-m_2,n,\ell)$.}
\label{table1}
\end{center}
\end{table}
Now, we will apply  the modified RT method  and achieve a closed form expression for the  HOMFLY-PT  polynomials of the knots of type $\hat{Q}_3(m_1,-m_2,n,\ell)$.

\subsection{HOMFLY-PT for twisted generalized hybrid weaving knots }
The HOMFLY-PT polynomial corresponds to  the fundamental representation ($[r]=[1]$) on each strand of the  braid. The tensor product decomposition: $$[1]^{\bigotimes 3}=[3,0,0]\bigoplus[1,1,1]\bigoplus 2 [2,1,0]$$ shows that the representation $[2,1,0]$ is of  multiplicity two.
 From  Equations \ref{HOMFLY}, \ref{R1} and \ref{R21}, the HOMFLY-PT for $\hat{Q}_3(m_1,-m_2,n,\ell)$ is given by the following formula.
\begin{eqnarray}
\mathcal{H}_{[1]}^{\hat{Q}_3(m_1,-m_2,n,\ell)} &=& \frac{1}{S^*_{[1]}}\sum_{\alpha} {\rm Tr}_{\Xi_{\alpha}}  (\hat{\mathfrak{R}}^{\Xi_{\alpha}}_1)^{m_1}(\hat{\mathfrak{R}}^{\Xi_{\alpha}}_2)^{-m_2})^n(\hat{\mathfrak{R}}^{\Xi_{\alpha}}_1\hat{\mathfrak{R}}^{\Xi_{\alpha}}_2)^{3\ell}~,\nonumber\\&=& \frac{1}{S^*_{[1]}} \Bigg(q^{n (m_1-m_2)+ 6 \ell}S^*_{[3,0,0]}  +(-q)^{-n(m_1-m_2)- 6 \ell} S^*_{[1,1,1]} +S^*_{[2,1,0]}\nonumber\\&&{\rm Tr}_{[2,1,0]}\Psi^{[2,1,0]}((m_1,-m_2,n,\ell);q)\Bigg){\label{HW11}}.
\end{eqnarray}
To obtain an explicit  formula for the  HOMFLYPT polynomial of  $\hat{Q}_3(m_1,-m_2,n, \ell)$, we need to evaluate the trace of the matrix $\Psi^{[2,1,0]}((m_1,-m_2,n,\ell);q)$. Since we have $$\Psi^{[2,1,0]}((m_1,-m_2,n,\ell);q)=((\hat{\mathfrak{R}}^{[2,1,0]}_1)^{m_1}(\hat{\mathfrak{R}}^{[2,1,0]}_2)^{-m_2}))^n(\hat{\mathfrak{R}}^{[2,1,0]}_1\hat{\mathfrak{R}}^{[2,1,0]}_2)^{3\ell},$$
We can use Equation \ref{R12} and Equation  \ref{R2} to show that
\begin{eqnarray*}
(\hat{\mathfrak{R}}^{[2,1,0]}_1)^{m_1}(\hat{\mathfrak{R}}^{[2,1,0]}_2)^{-m_2}&=& {A}^{-\alpha}\left(\begin{array}{cc} h_1(m_1 , m_2) & h_2(m_1 , m_2)\\ \\  \frac{h_2(m_1 , m_2)}{(-1)^{m_1}q^{2m_1}} & h_3(m_1 , m_2) \end{array}\right), \mbox{ and} \\
 (\hat{\mathfrak{R}}^{[2,1,0]}_1\hat{\mathfrak{R}}^{[2,1,0]}_2)^{3l}
&=& {A}^{-\beta}\left(\begin{array}{cc} 1 & 0\\ \\  0 & 1 \end{array}\right),\end{eqnarray*}~
where
\begin{eqnarray}
h_1(m_1 , m_2)&=&\frac{q^{\alpha } \left(1+(-1)^{m_2}q^{2 \text{$m_2$} }\text{}[3]_q\right)}{\text{}([2]_q)^2},~
h_2(m_1 , m_2)=\frac{q^{\alpha} \left(-1+(-1)^{m_2}q^{2 \text{$m_2$} }\right) \sqrt{\text{}[3]_q}}{\text{}([2]_q)^2},\nonumber\\
h_3(m_1 , m_2)&=&(-1)^{\alpha}\frac{q^{-\alpha } \left(1+(-1)^{-m_2}q^{-2 \text{$m_2$} } \text{}[3]_q\right)}{\text{}([2]_q)^2},~\alpha=(m_1-m_2),~\beta=6\ell. \nonumber
\end{eqnarray}
Note that as we assumed earlier,  $\alpha=(m_1-m_2)\geq 0$.  An  explicit form of $n^{th}$ power of the above matrix, denoted hereafter as $(\Psi^{[2,1,0]}((m_1,-m_2,n,\ell);A,q)$) is given by the following proposition. \\
{\bf Proposition 1.}
$$
\Psi^{[2,1,0]}(m_1,-m_2,n,l);A,q) =
{A}^{-\gamma}\left(\begin{array}{cc}\psi_{11}((m_1,-m_2,n);q) & \psi_{12}((m_1,-m_2,n-1);q) \\ \\\ \frac{\psi_{12}((m_1,-m_2,n-1);q)}{(-1)^{m_1}q^{2m_1}} & \psi_{22}((m_1,-m_2,n);q) \end{array}\right),
$$

where $\gamma=n \alpha+6\ell$, $\alpha=m_1-m_2$, and
\begin{eqnarray*}
\psi_{11}((m_1,-m_2,n);q) &=&\sum_{j=0
}^{\floor{\frac{n}{2}}}\sum_{k=1}^{n-j+1}(-1)^{-(m_1 j)} \binom{k+j-2}{j-1} \binom{n-(k+j-1)}{j}\\&&(h_1(m_1 , m_2)
)^{n-(2j+k-1)} (h_3(m_1 , m_2))^{k-1}(\frac{h_2(m_1 , m_2)}{q^{m_1}})^{2j},\\
\psi_{22}((m_1,-m_2,n);q)&=&\sum_{j=0
}^{\floor{\frac{n}{2}}}\sum_{k=1}^{n-j+1}(-1)^{-(m_1 j)} \binom{k+j-2}{j-1} \binom{n-(k+j-1)}{j} \\&&(h_3(m_1 , m_2))^{n-(2j+k-1)}(h_1(m_1 , m_2))^{k-1}(\frac{h_2(m_1 , m_2)}{q^{m_1}})^{2j},\\
\psi_{12}((m_1,-m_2,n);q)&=&\sum_{j=0}^{\floor{\frac{n}{2}}}\sum_{k=1}^{n-j+1}(-1)^{-(m_1 j)} \binom{k+j-1}{j} \binom{n-(k+j)}{j}\\&&(h_1(m_1 , m_2))^{n-(2j+k-1)} (h_3(m_1 , m_2))^{k-1}\frac{(h_2(m_1 , m_2))^{2j+1}}{q^{2j m_1}}.
\end{eqnarray*}
Notice that the entries in the matrix above are independent of the twist factor $\ell$ and the matrix structure is similar to the case of weaving knots $W(3,n)$, i.e.,  $\hat{Q}_3(1,-1,n,0)$ discussed in the parallel work \cite{RRV}. Hence, the proof of the proposition can be done in the same way as for   weaving knots. The trace of the matrix $(\Psi^{[2,1,0]}((m_1,-m_2,n,\ell);A,q)$ is given by
\begin{eqnarray}{\label{tt}}
\Psi^{[2,1,0]}((m_1,-m_2,n,\ell);A,q)&=&A^{-\gamma}(\psi_{11}((m_1,-m_2,n);q)+\psi_{22}((m_1,-m_2,n);q)),\nonumber\\&=&A^{-\gamma}\bigg(\sum_{j=0
}^{\floor{\frac{n}{2}}}\sum_{k=1}^{n-j}(-1)^{-(m_1 j )} \binom{k+j-2}{j-1}\binom{n-(k+j-1)}{j} \nonumber\\&&\big(h_1(m_1 , m_2))^{(n+1-2j-k)}( h_3(m_1 , m_2))^{(k-1)}+\nonumber\\&& (h_1(m_1 , m_2))^{(k-1)}(h_3(m_1 , m_2))^{(n-2j-k+1)}\big)\nonumber\\&&(\frac{h_2(m_1 , m_2)}{q^{m_1}})^{2j}\bigg).
\end{eqnarray}
Substituting the parameters in Equation \ref{tt}, we obtain the following expression
 \begin{eqnarray}{\label{Tracemod}}
\Psi^{[2,1,0]}((m_1,-m_2,n,\ell);A,q)&=&A^{-\gamma}\Omega((m_1,m_2,n);q),\\
\Omega((m_1,m_2,n);q)&=&(-1)^{\frac{-n\alpha}{2}}\sum_{j=0
}^{\floor{\frac{n}{2}}}\sum_{k=1}^{n-j}\binom{k+j-2}{j-1}
 \binom{n-(k+j-1)}{j}\nonumber\\&&G_{m_1,-m_2, j,k,n},\nonumber\\&&\text{where}\nonumber\\G_{m_1,-m_2, j,k,n}&=&\sum_{k_1=0}^{n+1-2j-k}\sum_{k_2=0}^{k-1}\sum_{k_3=0}^{2j}\sum_{\ell_1=1}^{2}\binom{n-(k+2j-1)}{k_1}\binom{2j}{k_3}\binom{k-1}{k_2} \nonumber\\&& \frac{{[3]_q}^{j+k_1+k_2}}{{[2]_q}^{2n}}(-1)^{\Big(\mu(\ell_1)(\alpha(k-1-\frac{n}{2})+m_2(k_2-k_1))+k_3(m_2+1)-m_1 j\Big)}\nonumber\\&& q^{\Big(\mu(\ell_1)(\alpha (-2 + 2 j + 2 k - n) +2 m_2 (k_2 - k_1))+2 m_2 (k_3 - j)\Big)}
 \nonumber,
\end{eqnarray}
here $\gamma=n \alpha+6\ell$, $\alpha=(m_1-m_2)$ and $\mu(\ell_1)$ denotes the Mobius function.

\begin{corollary}{\label{col}}
If  $m_1=m_2=m$, then the trace term reduces to
 \begin{eqnarray}{\label{Tracemod1}}
(\Psi^{[2,1,0]}((m,-m,n,\ell);A,q)&=&A^{-6\ell}\Omega((m,m,n);q) ~~~~~\text{where},\nonumber\\
\Omega((m,m,n);q)&=&\sum_{j=0
}^{\floor{\frac{n}{2}}}\sum_{k=1}^{n-j}(-1)^{-m j}\binom{k+j-2}{j-1}
 \binom{n-(k+j-1)}{j}\nonumber\\&&G_{m,-m, j,k,n},\\&&\text{and}\nonumber\\G_{m,-m, j,k,n}&=&\sum_{k_1=0}^{n+1-2j-k}\sum_{k_2=0}^{k-1}\sum_{k_3=0}^{2j}\sum_{\ell_1=1}^{2}(-1)^{\Big((\mu(\ell_1)m (k_2-k_1))+k_3(m+1)\Big)}\nonumber\\&&\binom{n-(k+2j-1)}{k_1}\binom{2j}{k_3}  \binom{k-1}{k_2}\frac{{[3]_q}^{j+k_1+k_2}}{{[2]_q}^{2n}}\nonumber\\&&q^{2m\Big(\mu(\ell_1) (k_2 - k_1)+(k_3-j)\Big)}\nonumber.
\end{eqnarray}
\end{corollary}
The trace $\Omega((m,m,n);q)$ is a symmetric Laurent polynomial in the variable $q$. We propose the following conjecture.\\
{\bf Conjecture 1.} {\it Let $m$ be a natural number. Then the sum of the absolute coefficients of the Laurent polynomial  $\Omega((m,m,n);q)$ is equal to the  $m^{th}$ Lucas number. In other words \\
 \begin{eqnarray}{\label{RLucas}}
\Omega((m,m,n);\iu)&=& {L}_{m,2n},\end{eqnarray}
where the symbol ${L}_{m,2n}$ denotes the  $m^{th}$ Lucas number generated by the  $m^{th}$ Fibonacci sequence \cite{Falcn2011}} and $\iu$ is the imaginary unit.\\

For clarity, we shall  briefly recall the definition of  the $m^{th}$  Fibonacci numbers  $F_{m,n}$ and the  $m^{th}$ Lucas numbers $L_{m,n}$. These  are sequences ($G_{m,n}$)  satisfying the Fibonacci recursive relation
\begin{eqnarray*}
G_{m,n+1}=m G_{m,n}+G_{m,n-1},
\end{eqnarray*}
 where $n\geq1$ and the  initial conditions are  $F_{m,0}=0, F_{m,1}=1$, $L_{m,0}=2$, and  $L_{m,1}=m$.  The special cases at $m=1$, $F_{1,n}$ and $L_{1,n}$ are called  classical Fibonacci and Lucas number respectively. They satisfy  the following  identities:
\begin{eqnarray*}
{L}_{m,n}&=&{F}_{m, n-1}+{F}_{m, n+1},~\text{for}~ n\geq1,\\
{L}^2_{m,n}&=&(m^2+4){F}^2_{m, n}+4(-1)^{n},\\
{L}^2_{m,n}+{L}^2_{m,n+1}&=&(m^2+4){F}_{m, 2n+1},\\
{L}_{m,2n}&=&{L}_{m, n}^2+2(-1)^{n+1}.
\end{eqnarray*}
The Binet Formula for $m^{th}$ Fibonacci number $F_{m,n}$ writes as
 \begin{eqnarray}{\label{Binet}}
F_{m,n}&=&\frac{\Phi_m^n-\Phi_m^{-n}}{\Phi_m+\Phi_m^{-1}},~\text{where}~~~ \Phi_m=\frac{m+\sqrt{m^2+4}}{2}.
\end{eqnarray}
Moreover, by Theorem 2.7 of \cite{Falcn2011},  we have  the  following  formula:   $$L_{m,2n}=\frac{1}{2^{2n-1}}\sum_{j=0}^{n}\binom{2n}{2j}m^{2n-2j}(m^2+4)^{j}.$$
We have tabulated the first terms  of the sequence  $L_{m,2n}$ for few values of $m$ in Table \ref{table3}.
\begin{table}[ht]
\begin{center}\small{
\begin{tabular} { | p {2 cm}| p {11cm}|}
\hline
$L_{m,2n}$&  $m^{th}$ \text{Lucas sequences for} $n \geq1 $ \\
\hline
$L_{1,2n}$ &$3, 7, 18, 47, 123, 322, 843, 2207, 5778, 15127, 39603, 103682,\ldots$\\
\hline
$L_{2,2n}$ &$6, 34, 198, 1154, 6726, 39202, 228486, 1331714, 7761798, 45239074,\ldots$\\
\hline
$L_{3,2n}$ &$11, 119, 1298, 14159, 154451, 1684802, 18378371, 200477279, \ldots$\\
\hline
$L_{4,2n}$ &$18, 322, 5778, 103682, 1860498, 33385282, 599074578, 10749957122,\ldots$\\
\hline
$L_{5,2n}$ &$27, 727, 19602, 528527, 14250627, 384238402, 10360186227,\ldots$\\
\hline
$L_{6,2n}$ &$38, 1442, 54758, 2079362, 78960998, 2998438562,\ldots$\\
\hline
$L_{7,2n}$ &$51, 2599, 132498, 6754799, 344362251, 17555720002,\ldots$\\
\hline
\end{tabular}}
\caption{The sequence $L_{m,2n}$ for small values of $m$.}
\label{table3}
\end{center}
\end{table}

Further, $\Phi_m$ satisfies  the following equation  $$ \Phi_m^2-m\Phi_m-1=0.$$
Let us now  test our conjecture Equation \ref{RLucas} in  the special  case $m=1,\ell=0$.  The trace term reduces to
\begin{eqnarray}{\label{Traceweav}}
\Omega((1,1,n);\iu)&=&L_{1,2n}\nonumber.
\end{eqnarray}
The result  coincides with what is  obtained in parallel work \cite{RRV}. Using Equation  \ref{Tracemod}, binomial series for the trace,  we get  the following:\\
{\bf Proposition 2.} The closed form expression of normalized two-variable HOMFLY-PT polynomial  for twisted generalized hybrid weaving knots $\mathcal{H}_{[1]}^{\hat{Q}_3(m_1,-m_2,n,\ell)} [A,q]$ turns out to be
\begin{eqnarray}{\label{HWK}}
\mathcal{H}_{[1]}^{\hat{Q}_3(m_1,-m_2,n,\ell)}&=&\frac{1}{S^*_{[1]} }\Bigg(\bigg(\frac{q}{A}\bigg)^{n \alpha+ 6 \ell}S^*_{[3,0,0]}  +(-q A)^{-n \alpha- 6 \ell} S^*_{[1,1,1]} + \bigg(\frac{1}{A}\bigg)^{n \alpha+6\ell}S^*_{[2,1,0]}\nonumber\\&&\cdot \Big( \sum_{j=0
}^{\floor{\frac{n}{2}}}\sum_{k=1}^{n-j}(-1)^{\frac{-n \alpha}{2}}\binom{k+j-2}{j-1}
 \binom{n-(k+j-1)}{i}G_{m_1,-m_2, j,k,n}\Big)\Bigg).\nonumber\\
\end{eqnarray}
Here  \begin{eqnarray}
G_{m_1,-m_2, j,k,n}&=&\sum_{k_1=0}^{n+1-2j-k}\sum_{k_2=0}^{k-1}\sum_{k_3=0}^{2j}\sum_{\ell_1=1}^{2}\binom{n-(k+2j-1)}{k_1}\binom{2j}{k_3}\binom{k-1}{k_2} \nonumber\\&& \frac{{[3]_q}^{j+k_1+k_2}}{{[2]_q}^{2n}}(-1)^{\Big(\mu(\ell_1)(\alpha(k-1-\frac{n}{2})+m_2(k_2-k_1))+k_3(m_2+1)-m_1 j\Big)}\nonumber\\&& q^{\Big(\mu(\ell_1)(\alpha (-2 + 2 j + 2 k - n) +2 m_2 (k_2 - k_1))+2 m_2 (k_3 - j)\Big)}\nonumber,
\end{eqnarray} and $\alpha=(m_1-m_2)$, $S^{*}_{[1]}=[N]_q$, $S^{*}_{[3,0,0]}=\frac{[N]_{q}[N+1]_{q}[N+2]_{q}}{[2]_{q}[3]_{q}}$, $S^{*}_{[1,1,1]}=\frac{[N]_{q}[N-1]_{q}[N-2]_{q}}{[2]_{q}[3]_{q}}$, $S^{*}_{[2,1,0]}=\frac{[N]_{q}[N+1]_{q}[N-1]_{q}}{[3]_{q}}$.\\

{\bf Examples.} If we apply the formula above to the knots  $\hat{Q}_3(1,-1,4,-1)={\bf 10_{157}}$ and $\hat{Q}_3(5,-1,1,-1)={\bf 8_{20}}$, we get:
\begin{eqnarray*}
H^{{\bf 10_{157}}}_{[1]}(q,A)&=&\frac{A^4}{q^6} (-A^2 + 2 q^2 + 3 A^2 q^2 + A^4 q^2 - 3 q^4 -
   5 A^2 q^4 - 3 A^4 q^4 + 4 q^6 +\\&& 6 A^2 q^6 + 3 A^4 q^6 - 3 q^8 -
   5 A^2 q^8 - 3 A^4 q^8 + 2 q^{10} + 3 A^2 q^{10} + A^4 q^{10} \nonumber\\&&- A^2 q^{12}).\\
H^{{\bf 8_{20}}}_{[1]}(q,A)&=&1 + 2 A^2 + \frac{A^2}{q^4} - \frac{1}{q^2} -\frac{ A^4}{q^2} - q^2 - A^4 q^2 + A^2 q^4.
\end{eqnarray*}
Substituting  $N=2$ in the formula of HOMFLY-PT in Equation \ref{HWK}, we obtain  the Jones polynomial of the knot $\hat{Q}_3(m_1,-m_2,n,\ell)$
\begin{eqnarray*}
\mathcal{J}^{\hat{Q}_3(m_1,-m_2,n,\ell)}(q) = (q^{n (m_1-m_2)+ 6\ell+2}+(q)^{n (m_1-m_2)+ 6\ell-2}+ \Psi^{[2,1,0]}((m_1,-m_2,n,\ell);q^2,q),\nonumber
\end{eqnarray*}
where the explicit form of $\Psi^{[2,1,0]}((m_1,-m_2,n,\ell);q^2,q)$ is as given in Equation  \ref{Tracemod}.
\begin{corollary}
The Jones polynomial of the  quasi-alternating knot $\hat{Q}_3(1,-1,n,\pm1)$ for $n\geq4$ is  given by
\begin{eqnarray*}
\mathcal{J}^{\hat{Q}_3(1,-1,n,\pm1)}(q) =  \sum_{k=-n}^{n}(-1)^{k}\hat{J}_{(n,k,\pm1)} q^{2k},\end{eqnarray*}
where the explicit form of $\hat{J}_{(n,k,\pm1)}
$ is
\begin{eqnarray}{\label{Traceweav-1}}
\hat{J}_{(n,k,\pm1)}&=&n \sum_{i=0}^{\floor{\frac{(n-\abs{k})}{2}}}  \frac{1}{n-i}\binom{n-i}{\abs{k}+i} \binom{n-\abs{k}-i-1}{i}+\delta_{k,\pm2}+\delta_{k,\pm4}.
\end{eqnarray}
\end{corollary}
Similarly, the Alexander polynomial $\Delta(q)$ of ${\hat{Q}_3(m_1,-m_2,n,\ell)}$ is  a specialization of the  HOMFLY-PT polynomial, obtained by setting  $A=1$.
\begin{eqnarray}{\label{alexa0}}
\Delta^{\hat{Q}_3(m_1,-m_2,n,\ell)}(q) ={\frac{1}{[3]_q}} (q^{n (m_1-m_2)+6\ell}+(-q)^{-n (m_1-m_2)-6\ell}-(\Psi^{[2,1,0]}((m_1,-m_2,n,\ell);1,q).\nonumber
\end{eqnarray}
Furthermore, if  $m_1= m_2=1$, we can neatly rewrite the above expression of Alexander polynomial of the knot  ${\hat{Q}_3(1,-1,n,\ell)}$ as follows
\begin{eqnarray}{\label{alexa}}
\Delta^{\hat{Q}_3(1,-1,n,\ell)}(q) =\Delta^{\hat{Q}_3(1,-1,n,0)}(q)+\sum_{-3l+2}^{3l}(-1)^{i-1}q^{2 (i-1)},
\end{eqnarray}
where $\Delta^{\hat{Q}_3(1,-1,n,0)}(q)$ is the Alexander polynomial of  weaving knots discussed in \cite{RRV}.
\begin{corollary}\label{corAlexQA}
The Alexander polynomial of the quasi-alternating knot ${\hat{Q}_3(1,-1,n,\pm 1)}$ for $n\geq3$ is given by
\begin{eqnarray}{\label{qalexa}}
\Delta^{\hat{Q}_3(1,-1,n,\pm 1)}(q) =\sum^{n-1}_{i=-(n-1)} \hat{\Delta}_{(n,\abs{i})}
 q^{2i},
\end{eqnarray}
where
\begin{eqnarray*}
\hat{\Delta}_{(n,i)}&=& \sum^{\floor{\frac{2n+2+i}{3}}}_{j=i+1} (-1)^{\floor{\frac{j+i-1}{2}}} (\phi[n,\floor{\frac{3j-i-1}{2}}]-\delta_{{i},1}+\delta_{i,2}), \\
\text{where}&&\\
\phi[n,k]&=& n \sum_{i=0}^{\floor{\frac{(n-\abs{k})}{2}}}  \frac{1}{n-i}\binom{n-i}{\abs{k}+i} \binom{n-\abs{k}-i-1}{i}.
\end{eqnarray*}
\end{corollary}
For clarity, we have listed $\phi[n,k]$ for few values of $k$ in Table \ref{table2}.
\begin{table}[ht]
\centering
\small{
\begin{tabular} { | p {2cm}| p {.6 cm}| p {1 cm}| p {2 cm}| p {3 cm}| p {4 cm}|}
\hline
$k$ &0&1& 2& 3& 4\\
\hline
$\phi[n,n-k]$&1 &$n$&$\frac{1}{2}(n-1)n$&$\frac{1}{6}n(n^2-3n+8)$&$\frac{1}{24}n(n-2)(n^2-4n+27)$\\
\hline
\end{tabular}}
\caption{ $\phi[n,n-k]$ for $ k \leq 4$. }
\label{table2}
\end{table}

{\bf Remark 1.} {\it The knot ${\hat{Q}_3(1,-1,n,\pm 1)}$ is quasi-alternating. Consequently,  its link Floer homology is thin. More precisely,  this homology is determined by  the coefficients of the Alexander polynomial and the knot signature. Note that the signature of ${\hat{Q}_3(1,-1,n,\pm 1)}$ is $\pm 4$. Thus, Corollary \ref{corAlexQA} can be used  also to find  the ranks of the link Floer homology of the knot ${\hat{Q}_3(1,-1,n,\pm 1)}$.}

\subsection{The determinants of twisted hybrid weaving knots}
Recall that the determinant of an oriented link $L$, ${\rm{Det}}(L)$ is a numerical invariant of links that is defined from the Seifert matrix of the link. This invariant can be obtained as the evaluation of the Alexander, or the Jones  polynomial at -1, i.e.,  ${\rm{Det}}(L)=|\Delta^{L}(-1)|=|\mathcal{J}^L(-1)|$.
The determinants of 3-braid links has been studied in \cite{QC}. The calculation in the previous section reveals a connection between  the determinant of the knot  $\hat{Q}_3(m,-m,n,\ell)$, where $m$ is
a natural number, and the  $m^{th}$  Lucas number $L_{m,2n}$. We suggest the following conjecture.\\
{\bf Conjecture 2.} {\it Given  $m$ natural number and $\ell\in\{-1,0,1\}$, then we have:
 \begin{eqnarray}{\label{CHP}}
{\rm Det}(\hat{Q}_3(m,-m,n,0))&=&|\Delta^{\hat{Q}_3(m,-m,n,0)}(-1)|= L_{m,2n} -2,\nonumber\\
{\rm Det}(\hat{Q}_3(m,-m,n,\pm 1))&=&|\Delta^{\hat{Q}_3(m,-m,n,\pm1)}(-1)|= L_{m,2n}+2,
~\end{eqnarray}
where $L_{m,2n}$ is $m^{th}$ Lucas number. }\\
{\bf Remark 2.} {\it
The classical Lucas number  and golden ratio correspond to the case   $m=1$. In other words, $L_{1,2n}=L_{2n}$ and $\Phi_1=\frac{1 + \sqrt{ 5}}{2}$. It has been proved in \cite{Layla, Stees, RRV} that for weaving knot
 $Det(W (3,n))=L_{2n}-2$.}\\
The closed form expression obtained above for the Alexander polynomial   is a   good starting point to investigate the  trapezoidal behavior of the Alexander polynomial of alternating closed 3-braids.
\subsection{Trapezoidal Conjecture for alternating knots}

In 1962, R. Fox conjectured that the coefficients of the Alexander polynomial of alternating knots are trapezoidal \cite{Fox}. In other words, the absolute values of these coefficients increase, stabilize than decrease in a symmetrical way. This conjecture, known as  Fox’s trapezoidal conjecture, has been confirmed
for several classes of alternating knots. In particular, for some families of alternating three-braids, see \cite{AC}. We shall now prove that this conjecture holds for knots of type ${\hat{Q}_3(1,-1,n,\ell)}$, where   $\ell=0$ or $\ell=\pm 1$.\\

\begin{figure}[h]
\begin{center}$
\begin{array}{lll}
\includegraphics[width=75mm]{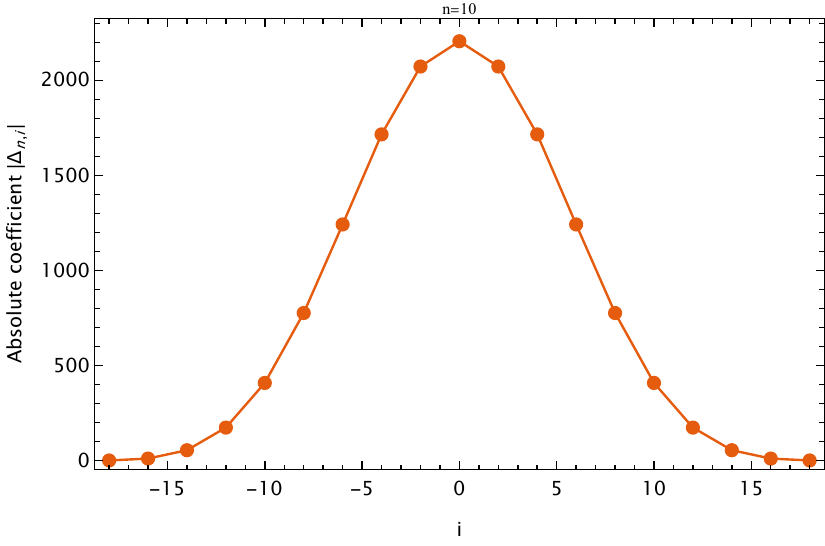}&
\includegraphics[width=75mm]{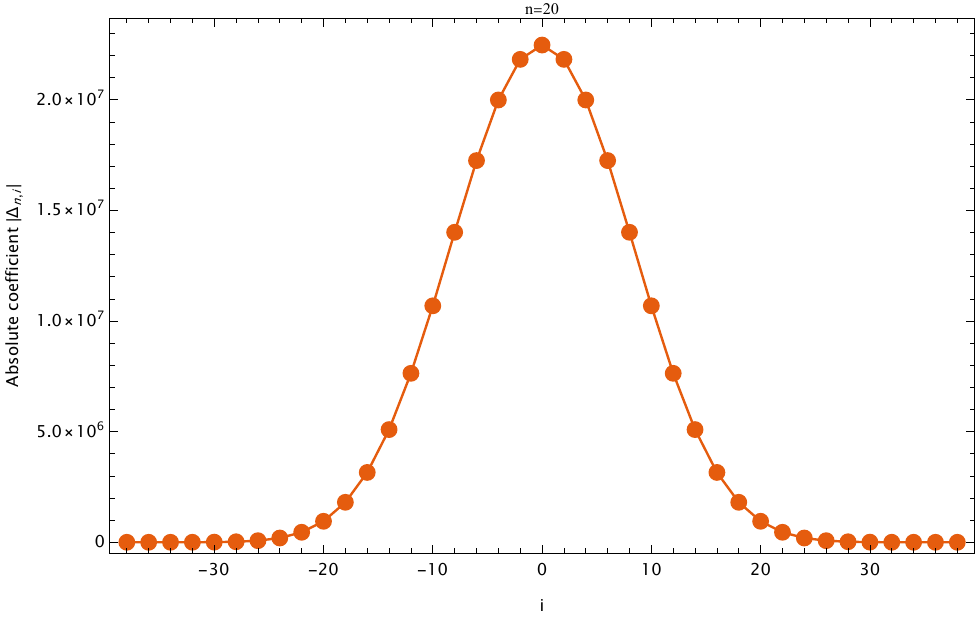}
\end{array}$
\end{center}
\begin{center}$
\begin{array}{rr}
\includegraphics[width=75mm]{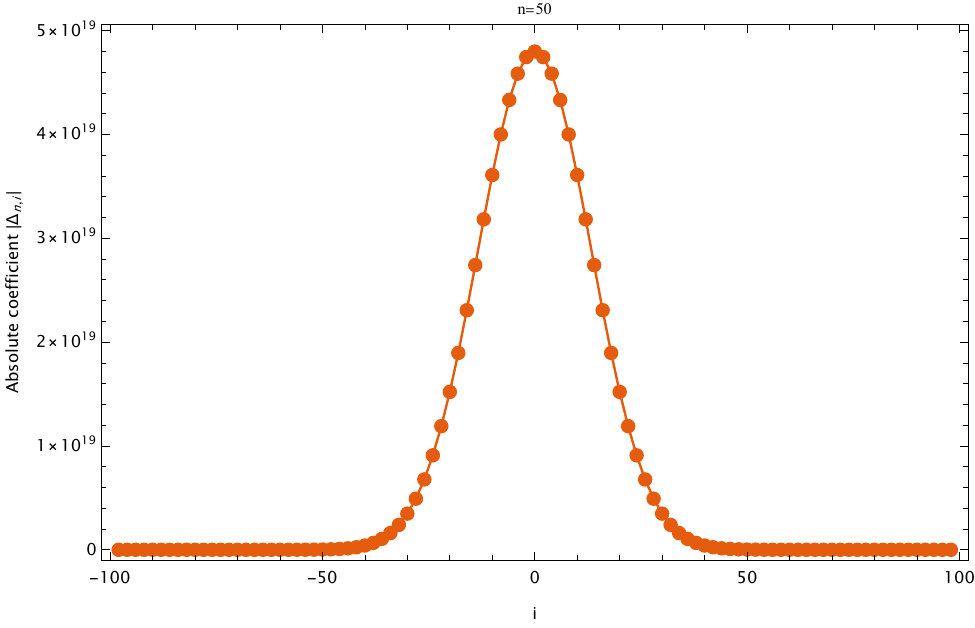}&
\includegraphics[width=75mm]{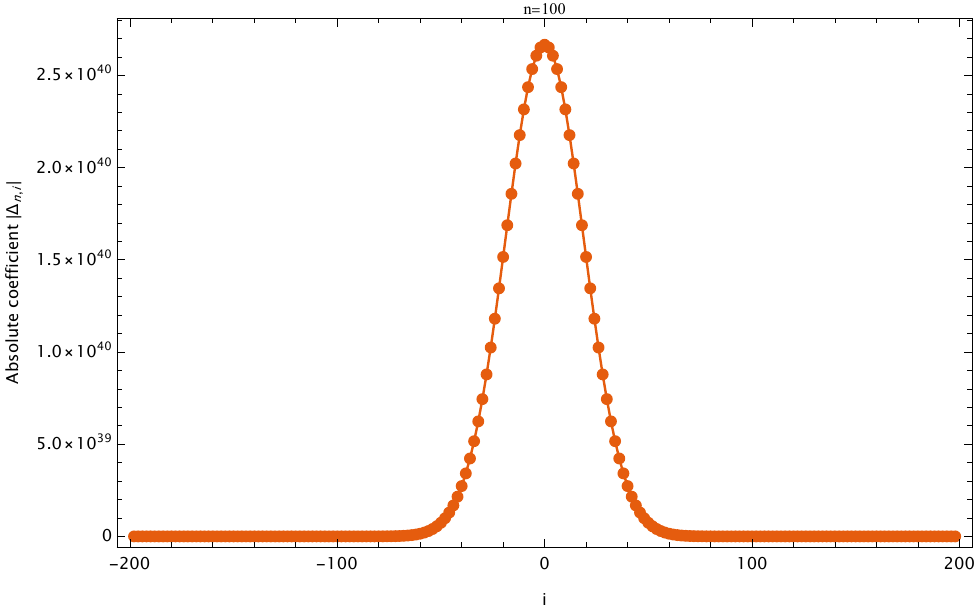}
\end{array}$
\end{center}
\caption{The distribution of absolute values of the  coefficients of Alexander polynomials of  $\hat{Q}_3(1,-1,n,\pm1)$.}
\label{TP}
\end{figure}

The Alexander polynomial of  knots of type ${\hat{Q}_3(1,-1,n,\ell)}$ in Equation \ref{alexa} is  given by
\begin{eqnarray}{\label{alexa2}}
\Delta^{\hat{Q}_3(1,-1,n,\ell)}(q) &=&\Delta^{\hat{Q}_3(1,-1,n,0)}(q)+\sum_{-3l+2}^{3l}(-1)^{i-1}q^{2 (i-1)},
\end{eqnarray}
where $\Delta^{\hat{Q}_3(1,-1,n,0)}(q)$ is the Alexander polynomial of the weaving knot in \cite{RRV}.
Numerically, we have checked the trapezoidal conjecture for large values of $n$ and $\ell=\pm 1$ for quasi-alternating knots, see  Figure \ref{TP}.

 Further, for large values of $n>>\ell$, we can prove the  following.
\begin{theorem} The asymptotic nature of the absolute values of the  coefficients of the  Alexander polynomial of the  knot $\hat{Q}_3(1,-1,n,\ell)$ for $n>>\ell$ is trapezoidal, i.e.,
\begin{eqnarray}{\label{trapo}}
\Delta^{\hat{Q}_3(1,-1,n,\ell)}(q)&\equiv&\Delta^{\hat{Q}_3(1,-1,n,0)}(q) =\sum^{n-1}_{i=-(n-1)} \mu_{(n,i)}
 q^{2i},~~~ n>>\ell,
\end{eqnarray}
where,
\begin{eqnarray*}
\delta_{(n,i)}&=&\abs{\mu_{(n,i)}}-\abs{\mu_{(n,i+1)}}\geq 0.
\end{eqnarray*}
\end{theorem}
\noindent {Proof.}  Using Equation \ref{alexa2}, we can rearrange Equation \ref{trapo} such that the difference will be
\begin{eqnarray}{\label{trapo1}}
\delta_{(n,0)}&=&\abs{\mu_{(n,0)}}-\abs{\mu_{(n,1)}}= \sum^{\floor{\frac{(2n+1)}{3}}}_{i=1}(-1)^{(\floor{\frac{i - 1}{2}} + i - 1)} \left(\phi[n,\floor{\frac{i}{2}}+ i]\right).~\end{eqnarray}

It is clear  from  Table \ref{table2} that
\begin{eqnarray}\lim_{n\to \infty}\phi[n,k]\approx	\frac{ n^{n-k}}{k!}\end{eqnarray}
From  Equation  \ref{trapo1},  we have
\begin{eqnarray*}
\delta_{(n,0)}&=& \sum^{\floor{\frac{(2n+1)}{3}}}_{i=1}(-1)^{(\floor{\frac{i - 1}{2}} + i - 1)} \left(\phi[n,\floor{\frac{i}{2}}+ i]\right),\\
&=&\phi[n,1]-\phi[n,3]-\phi[n,4]+\phi[n,6]+\phi[n,7]-\phi[n,9]-\phi[n,10]\ldots~,\\
&= &n^{n-1}-\frac{n^{n-3}}{6}-\frac{n^{n-4}}{24}+\frac{n^{n-6}}{720}+\frac{n^{n-7}}{5040}-\ldots~,\text{for large}~
 n,\\
&=&n^{n-1}(1-\frac{1}{6 n^{2}}-\frac{1}{24 n^{3}}+\frac{1}{720 n^{5}}+\frac{1}{5040 n^{6}}-\ldots).
\end{eqnarray*}
From the last equality above, we can see that $\delta_{(n,0)}>0$ for large values of $n$.  This completes the proof in this case. A similar argument  can be used  to prove that  $\delta_{(n,i)}\geq 0$ for any value of $i$.
\section{Colored HOMFLY-PT for twisted weaving  knots  $\hat{Q}_3(1,-1,n,\ell)$}\label{s4.int}
In the previous section, we  have derived an  explicit formula for the trace of  $2\otimes 2$ matrices (\ref{tt})  and HOMFLY-PT polynomial (\ref{HWK}) corresponding to the  representation $[r]=[1]$. In the following sub-section we shall discuss the generalization of these results to higher colors $[r]>1$.
\subsection{Trace terms for representations $[r]>1$}
 Given a representation $[r]>[1]$, the irreducible representation $\Xi_{\alpha}\equiv[{\xi_1}^{\alpha},{\xi_2}^{\alpha},{\xi_3}^{\alpha}]\in[r]^{\otimes 3}$ has multiplicity $(M)$ arising  from path structure of $\Lambda_{\alpha,\mu}\in [r]^{\otimes2}$. Here, $\mu$ takes the values $0,1,\ldots M-1$ and keeps track of multiplicity. Corresponding block diagonal matrix $\hat{\mathfrak{R}}^{\Xi_{\alpha}}_1$   entries are  $$\pm \{q^{\chi_{r,0}},-q^{\chi_{r,1}},\ldots (-1)^{M-1} q^{\chi_{r,M-1}}\},$$ and $\hat{\mathfrak{R}}^{\Xi_{\alpha}}_2=\mathcal{U}^{\Xi_{\alpha}}\hat{\mathfrak{R}}^{\Xi_{\alpha}}_1\mathcal{U}^{\Xi_{\alpha}}$.  After analyzing numerous examples of  twisted weaving  knots of braid index 3,  we propose the following:\\
  \textbf{Proposition 3.}
\begin{eqnarray}{\label{conj2}}
(\hat{\mathfrak{R}}_1^{\Xi_{\alpha}}\hat{\mathfrak{R}}_2^{\Xi_{\alpha}})^{3 \ell}&=&q^{\Lambda_M} \mathcal{I}_{M\times M},\nonumber\\
{\rm Tr}(\hat{\mathfrak{R}}_1^{\Xi_{\alpha}}(\hat{\mathfrak{R}}_2^{\Xi_{\alpha}})^{-1})^n(\hat{\mathfrak{R}}_1^{\Xi_{\alpha}}\hat{\mathfrak{R}}_2^{\Xi_{\alpha}})^{3 \ell}&=&q^{\Lambda_M}{\rm Tr}(\hat{\mathfrak{R}}_1^{\Xi_{\alpha}}(\hat{\mathfrak{R}}_2^{\Xi_{\alpha}})^{-1})^n,
\end{eqnarray}
here, ${\Lambda_M}$ is a  twist factor which depends on  parameters $\{\chi_{r,0},\chi_{r,1},\dots \chi_{r,M-1},\ell\}$ and $\mathcal{I}_{M \times M}$ is identity matrix of size $M$.\\
By checking several  examples, we compute  the twist factors for $M=2$ and $3$
 $$\Lambda_2={3(\chi_{r,0}+\chi_{r,1})\ell}, {\rm{and }}~~\Lambda_{3}={2(\chi_{r,0}+\chi_{r,1}+\chi_{r,2})\ell}.$$
At the moment it is difficult to guess a closed form of $\Lambda_{M}$ for $M \geq 4$.
Let $$\Theta_{n,t,m,\ell}[q]={\rm Tr}(\hat{\mathfrak{R}}_1^{\Xi_{\alpha}}(\hat{\mathfrak{R}}_2^{\Xi_{\alpha}})^{-1})^n(\hat{\mathfrak{R}}_1^{\Xi_{\alpha}}\hat{\mathfrak{R}}_2^{\Xi_{\alpha}})^{\ell}\nonumber.$$
The following  conjecture  generalizes  Proposition 1 in \cite{VRR}.\\
\textbf{Conjecture 3.} Given a representation $\Xi_{\alpha}\equiv[{\xi_1}^{\alpha},{\xi_2}^{\alpha},{\xi_3}^{\alpha}]$ having multiplicity 2  with  $\hat{\mathfrak{R}}^{\Xi_{\alpha}}_1=\pm \left(\begin{array}{cc} q^{t+m} & 0 \\ \\  0 & -q^{-t+m} \end{array}\right)$, $\hat{\mathfrak{R}}^{\Xi_{\alpha}}_2=\mathcal{U}^{\Xi_{\alpha}}\hat{\mathfrak{R}}^{\Xi_{\alpha}}_1\mathcal{U}^{\Xi_{\alpha}}$,~and~  $\mathcal{U}^{\Xi_{\alpha}}=\left(\begin{array}{cc}
\frac{1}{[2]_{q^t}} & \frac{\sqrt{[3]_{q^t}}}{ [2]_{q^t}} \\
\frac{\sqrt{[3]_{q^t}}}{[2]_{q^t}} & -\frac{1}{[2]_{q^t}}
\end{array}\right).\ \ $ Then
\begin{eqnarray}{\label{conj3}}
\Theta_{n,t,m,\ell}[q]&=&q^{\Lambda_2}{\rm Tr}(\hat{\mathfrak{R}}_1^{\Xi_{\alpha}}(\hat{\mathfrak{R}}_2^{\Xi_{\alpha}})^{-1})^n,\nonumber\\
&=&\sum_{g=-n}^{n} (-1)^{g} \Omega_{n,n-\abs{g}}q^{2 g t+6 m \ell}.
\end{eqnarray}
Here  $t$ and $m$ are   integers which  dependent on $\Xi_{\alpha}$ and the coefficients $\Omega_{n,j}$ are given by the following formula:
\begin{eqnarray*}
\Omega_{n,j}&=&\sum_{i=0}^{\floor{\frac{j}{2}}}\frac{n}{n-i}\binom{n-i}{n-j+ i} \binom{j-i-1}{i}~,
\end{eqnarray*}
where the parameters $n$ and $j$ are positive integers,  $|x|$ denotes the absolute value of $x$ and $\floor{x}$ indicates the greatest integer less than or equal to  $x$.\\
In the following subsections, we will
use  Equations  \ref{conj2} and \ref{conj3} to compute   the $[r]$-colored HOMFLY-PT polynomials  of  $\hat{Q}_3(1,-1,n,\ell)$ for $[r]=[2],[3], \text{and}~ [4]$.
\subsection{HOMFLY-PT polynomial of the knots $\hat{Q}_3(1,-1,n,\ell)$}
The HOMFLY-PT polynomial of  the  knot $\hat{Q}_3(1,-1,n,\ell)$ is given by the following formula:
\begin{eqnarray}{\label{WEAVING}}
\mathcal{H}_{[1]}^{\hat{Q}_3(1,-1,n,\ell)} (A,q)= \frac{1}{S^*_{[1]}}(q^{6\ell}S^*_{[3]}  +(-q)^{- 6\ell}) S^*_{[111]} + S^*_{[21]}\Theta_{n,1,0,\ell}[q] ).
\end{eqnarray}
Substituting  $A=q^2$, we get the Jones polynomial:
\begin{eqnarray*}
\mathcal{J}^{\hat{Q}_3(1,-1,n,\ell)}(q)=q^{-2+6\ell}+q^{2+6\ell}+\Theta_{n,1,0,\ell}[q] ~.
\end{eqnarray*}
We will use the data on $U^{\Xi_{\alpha}}$ matrices in Section~\ref{s12.int} for three-strand braids  where $\Xi_{\alpha} \in  [r]^3$ to compute the  $[r]$-colored HOMFLY-PT polynomials of  the twisted weaving knots $\hat{Q}_3(1,-1,n,\ell)$.

\subsection{$[2]$-colored HOMFLY-PT polynomial of the knots $\hat{Q}_3(1,-1,n,\ell)$\label{se.symrac1}}
In the case $[r]=[2]$, the tensor product decomposition rules are as follows:
$$\bigotimes^3 [2] = [6,0,0]\bigoplus [3,3,0]\bigoplus  [4,1,1]\bigoplus[2,2,2]\bigoplus 2 [5,1,0]\bigoplus 2 [3,2,1]\bigoplus 3 [4,2,0].$$\\
It can be easily seen that there are one $3\times 3$-matrix, one $2\times 2$-matrix and  two $1\times 1$-matrices.  The twisted trace (\ref{conj3}) and twist factor (\ref{conj2}) are shown in the Table  {\ref{table4}}.\\

{\small \begin{table}[h]
\begin{center}
\begin{tabular}{ |p{3cm}|p{2cm}|p{3cm}|p{3cm}|}
\hline
$\Xi_{\alpha} \in [2]^3$  & Matrix size (M)& Trace $\Theta_{n,t,m,\ell}[q]$& Twist factor $\Lambda_{M}$ \\
\hline
[3,2,1] &2  & $\Theta_{n,1,1,\ell}[q]$& $6 \ell$\\
\hline
[5,1,0] &2  & $\Theta_{n,2,4,\ell}[q]$& $24 \ell$\\
\hline
[4,2,0]& 3 & To be determined & $16 \ell$\\
\hline
\end{tabular}
\caption{Twisted trace and twisted  factor table for $\Xi_{\alpha} \in  [2]^{\otimes 3}$. }\label{table4}
\end{center}
\end{table}
}

The eigenvalues matrices in this case are
\begin{equation*}\label{u0}
\hat{\mathfrak{R}}^{[5,1,0]}_1=q^4\left(
\begin{array}{cc}
 q^2 & 0 \\
 0 & -q^{-2} \\
\end{array}
\right)\\,~~
\hat{\mathfrak{R}}^{[4,2,0]}_1=\left(
\begin{array}{ccc}
 1 & 0 & 0 \\
 0 & -q^2 & 0 \\
 0 & 0 & q^6 \\
\end{array}
\right),
\end{equation*}
~\text{and}~
\begin{equation}
\hat{\mathfrak{R}}^{[3,2,1]}_1=q \left(
\begin{array}{cc}
 q^{-1} & 0 \\
 0 & -q \\
\end{array}
\right),\ \
\end{equation}
and the corresponding  $U^{\Xi_{\alpha}}$ extracted from Equation  \ref{symrac}. From  Equation \ref{HOMFLY}, the  $[2]$-HOMFLY-PT for $\hat{Q}_3(1,-1,n,\ell)$ can be expressed as follows:
\begin{eqnarray}
\mathcal{H}_{[2]}^{\hat{Q}_3(1,-1,n,\ell)} &=& \frac{A^{-12\ell}q^{-24\ell}}{S^*_{[2]}} (S^*_{[2,2,2]}+q^{ 36 \ell}S^*_{[6]}+(-q)^{
 12\ell} S^*_{[3,3]} +(-q)^{ 12\ell} S^*_{[4,1,1]}+\nonumber\\&&S^*_{[5,1]}{\rm Tr}_{[5,1,0]}  (\hat{\mathfrak{R}}^{[5,1,0]}_1(\hat{\mathfrak{R}}^{[5,1,0]}_2)^{-1})^n(\hat{\mathfrak{R}}^{[5,1,0]}_1\hat{\mathfrak{R}}^{[5,1,0]}_2)^{3\ell}\nonumber\\&&+S^*_{[3,2,1]}{\rm Tr}_{[3,2,1]}  (\hat{\mathfrak{R}}^{[3,2,1]}_1\hat{\mathfrak{R}}^{[3,2,1]}_2)^{-1})^n(\hat{\mathfrak{R}}^{[3,2,1]}_1\hat{\mathfrak{R}}^{[3,2,1]}_2)^{3\ell}\nonumber\\&&+S^*_{[4,2,0]}{\rm Tr}_{[4,2,0]}  (\hat{\mathfrak{R}}^{[4,2,0]}(\hat{\mathfrak{R}}^{[4,2,0]}_2)^{-1})^n(\hat{\mathfrak{R}}^{[4,2,0]}_1\hat{\mathfrak{R}}^{[4,2,0]}_2)^{3\ell}{\label{HW221}}.
\end{eqnarray}
Using Table {\ref{table4}}, we can rewrite  Equation~(\ref{HW221}) into a more concise  formula
\begin{eqnarray}
\mathcal{H}_{[2]}^{\hat{Q}_3(1,-1,n,\ell)} &=& \frac{A^{-12\ell}q^{-24\ell}}{S^*_{[2]}} (S^*_{[2,2,2]}+q^{ 36 \ell}S^*_{[6,0,0]}+(-q)^{ 12\ell} S^*_{[4,1,1]}+(-q)^{ 12\ell}S^*_{[3,3,0]}\nonumber\\&&+S^*_{[5,1]} \Theta_{n,2,4,\ell}[q]+S^*_{[3,2,1]} \Theta_{n,1,1,\ell}[q]  +q^{ 16 \ell}S^*_{[4,2,0]}{\rm Tr} (X^{[4,2,0]})^n,\nonumber\\{\label{HW2}}~
\end{eqnarray}
where,
\begin{equation}
X^{[4,2,0]}=\left(
\begin{array}{ccc}
 \frac{1}{q^6+q^8+q^{10}} & -\frac{1}{q^5 \sqrt{1+q^2+q^4}} & \frac{\sqrt{1+q^2+q^4+q^6+q^8}}{q^2+q^4+q^6} \\
 \frac{1}{q^3 \sqrt{1+q^2+q^4}} & \frac{-1+q^2-q^4}{q^2+q^6} & -\frac{q^3 \sqrt{\frac{1+q^2+q^4+q^6+q^8}{1+q^2+q^4}}}{1+q^4} \\
 \frac{q^4 \sqrt{1+q^2+q^4+q^6+q^8}}{1+q^2+q^4} & \frac{q^7 \sqrt{\frac{1+q^2+q^4+q^6+q^8}{1+q^2+q^4}}}{1+q^4} & \frac{q^{14}}{1+q^2+2 q^4+q^6+q^8}
\\
\end{array}
\right).
\end{equation}
It is worth mentioning here that the $[2]$-colored polynomial in Equation \ref{HW221} for arbitrary $n$ and $\ell$ are easily computable. We have listed $[2]$-colored polynomials in Appendix~\ref{app1} for some twisted weaving knots.

\subsection{$[3]$-colored HOMFLY-PT polynomial of the knots $\hat{Q}_3(1,-1,n,\ell)$ \label{t.symrac1}}
The tensor product  decomposition rules in this case writes as follows: \begin{eqnarray*}\bigotimes^3 [3]& =& [9,0,0]\bigoplus [7,1,1]\bigoplus  [5,2,2]\bigoplus  [4,4,1]\bigoplus  [3,3,3]\bigoplus 3 [8,1,0]\\ &&\bigoplus
 2[4,3,2]\bigoplus 2 [6,2,1]\bigoplus 2 [5,4,0]\bigoplus 3 [7,2,0]\bigoplus 3 [5,3,1]\\ &&\bigoplus 4 [6,3,0].\end{eqnarray*}
Thus, we have two $3\times 3$-matrices, one $2\times 2$-matrix, one $1\times 1$-matrix and one $4\times 4$-matrix as tabulated below.
{\small \begin{table}[h]
\begin{center}
\begin{tabular}{ |p{3cm}|p{3cm}|p{3cm}|p{3cm}|}
\hline
$\Xi_{\alpha} \in [3]^3$  & Matrix size $(M)$& Trace $\Theta_{n,t,m,\ell}[q]$& Twist factor $\Lambda_{M,}$ \\
\hline
[4,3,2] &2  & $\Theta_{n,1,4,\ell}[q]$& ${24 \ell}$\\
\hline
[6,2,1] &2  & $\Theta_{n,2,7,\ell}[q]$& ${42 \ell}$\\
\hline
[5,4,0]& 2 & $\Theta_{n,2,7,\ell}[q]$& ${42 \ell}$\\
\hline
[8,1,0]& 2 &$ \Theta_{n,3,12,\ell}[q]$& ${72 \ell}$\\
\hline
[7,2,0]& 3 & To be determined & ${58\ell}$\\
\hline
[5,3,1]& 3 & To be determined & ${34 \ell}$\\
\hline
[6,3,0]& 4 & To be determined & ${48 \ell}$\\
\hline
\end{tabular}
\caption{Twisted trace and twisted factors table for $\Xi_{\alpha} \in  [3]^{\otimes 3}$. }\label{table5}
\end{center}
\end{table}
}\\

The braiding matrices in this case are

\begin{equation*}{\label{col3}}
~{\hat{\mathfrak{R}}^{[6,3,0]}=\left(
\begin{array}{cccc}
 -q^3 & 0 & 0 & 0 \\
 0 & q^5 & 0 & 0 \\
 0 & 0 & -q^9 & 0 \\
 0 & 0 & 0 & q^{15} \\
\end{array}
\right)}\\,~~{\hat{\mathfrak{R}}^{[5,3,1]}=\left(
\begin{array}{ccc}
 -q^3 & 0 & 0 \\
 0 & q^5 & 0 \\
 0 & 0 & -q^9 \\
\end{array}
\right)}\\,
\end{equation*}

\begin{equation*}{\label{col31}}
{\hat{\mathfrak{R}}^{[7,2,0]}=\left(
\begin{array}{ccc}
 q^5 & 0 & 0 \\
 0 & -q^9 & 0 \\
 0 & 0 & q^{15} \\
\end{array}
\right)}~,
\hat{\mathfrak{R}}^{[5,4,0]}=\hat{\mathfrak{R}}^{[6,2,1]}=q^7\left(
\begin{array}{cc}
 q^{-2} & 0 \\
 0 & -q^2 \\
\end{array}
\right)\\,~~
~\end{equation*}
\begin{equation*}
\hat{\mathfrak{R}}^{[4,3,2]}=q^4\left(
\begin{array}{cc}
 -q^{-1} & 0 \\
 0 & q \\
\end{array}
\right) \\,~{\hat{\mathfrak{R}}^{[8,1,0]}=q^{12}\left(
\begin{array}{cc}
 -q^{-3} & 0 \\
 0 & q^{3} \\
\end{array}
\right)},
\end{equation*}~
and $U^{\Xi_{\alpha}}$ matrices computed from Equation  \ref{symrac}.
Using the Equation \ref{HOMFLY}, the  $[3]$-colored  HOMFLY-PT polynomial of  $\hat{Q}_3(1,-1,n,\ell)$ is  as given below:
\begin{eqnarray}
\mathcal{H}_{[3]}^{\hat{Q}_3(1,-1,n,\ell)} &=& \frac{A^{-36\ell}q^{-72\ell}}{S^*_{[3]}}\sum_{\alpha} {\rm Tr}_{\Xi_{\alpha}}  (\hat{\mathfrak{R}}^{\Xi_{\alpha}}_1)(\hat{\mathfrak{R}}^{\Xi_{\alpha}}_2)^{-1})^n(\hat{\mathfrak{R}}^{\Xi_{\alpha}}_1\hat{\mathfrak{R}}^{\Xi_{\alpha}}_2)^{3\ell}~,\nonumber\\&=& \frac{1}{S^*_{[3]}} (q^{90\ell}S^*_{[9]}+(-q)^{54\ell}S^*_{[7,1,1]}+q^{30\ell}S^*_{[5,2,2]}+q^{30\ell}S^*_{[4,4,1]}+(-q)^{18\ell}\nonumber\\&&S^*_{[3,3,3]}+S^*_{[5,4,0]}{\rm Tr}_{[5,4,0]} (\hat{\mathfrak{R}}^{[5,4,0]}_1(\hat{\mathfrak{R}}^{[5,4,0]}_2)^{-1})^n(\hat{\mathfrak{R}}^{[5,4,0]}_1\hat{\mathfrak{R}}^{[5,4,0]}_2)^{3\ell}+\nonumber\\&&S^*_{[4,3,2]}{\rm Tr}_{[4,3,2]} (\hat{\mathfrak{R}}^{[4,3,2]}_1(\hat{\mathfrak{R}}^{[4,3,2]}_2)^{-1})^n(\hat{\mathfrak{R}}^{[4,3,2]}_1\hat{\mathfrak{R}}^{[4,3,2]}_2)^{3\ell}+S^*_{[6,2,1]}\nonumber\\&&{\rm Tr}_{[6,2,1]} (\hat{\mathfrak{R}}^{[6,2,1]}_1(\hat{\mathfrak{R}}^{[6,3,1]}_2)^{-1})^n(\hat{\mathfrak{R}}^{[6,2,1]}_1\hat{\mathfrak{R}}^{[6,2,1]}_2)^{3\ell}+ S^*_{[8,1,0]} \nonumber\\&&{\rm Tr}_{[8,1,0]} (\hat{\mathfrak{R}}^{[8,1,0]}_1(\hat{\mathfrak{R}}^{[8,1,0]}_2)^{-1})^n(\hat{\mathfrak{R}}^{[8,1,0]}_1\hat{\mathfrak{R}}^{[8,1,0]}_2)^{3\ell}
+S^*_{[7,2,0]}\nonumber\\&&{\rm Tr}_{[7,2,0]}(\hat{\mathfrak{R}}^{[7,2,0]}_1\hat{\mathfrak{R}}^{[7,2,0]}_2)^{-1})^n(\hat{\mathfrak{R}}^{[7,2,0]}_1\hat{\mathfrak{R}}^{[7,2,0]}_2)^{3\ell}+S^*_{[5,3,1]}\nonumber\\&&{\rm Tr}_{[5,3,1]}(\hat{\mathfrak{R}}^{[5,3,1]}_1\hat{\mathfrak{R}}^{[5,3,1]}_2)^{-1})^n(\hat{\mathfrak{R}}^{[5,3,1]}_1\hat{\mathfrak{R}}^{[5,3,1]}_2)^{3\ell}+S^*_{[6,3,0]}\nonumber\\&&{\rm Tr}_{[6,3,0]}  (\hat{\mathfrak{R}}^{[6,3,0]}_1(\hat{\mathfrak{R}}^{[6,3,0]}_2)^{-1})^n(\hat{\mathfrak{R}}^{[6,3,0]}_1\hat{\mathfrak{R}}^{[6,3,0]}_2)^{3\ell}{\label{HW3}}.
\end{eqnarray}
Using Table $\ref{table5}$, we can rewrite  Equation \ref{HW3} into the following  formula
\begin{eqnarray}
\mathcal{H}_{[3]}^{\hat{Q}_3(1,-1,n,\ell)} &=& \frac{A^{-18\ell}q^{-72\ell}}{S^*_{[3]}} (q^{90\ell}S^*_{[9]}+(-q)^{54\ell}S^*_{[7,1,1]}+q^{30\ell}S^*_{[5,2,2]}+q^{30\ell}S^*_{[4,4,1]}\nonumber\\&&+(-q)^{18\ell}S^*_{[3,3,3]}+S^*_{[5,4,0]} \Theta_{n,2,7,\ell}[q]+S^*_{[6,2,1]}\Theta_{n,2,7,\ell}[q]+S^*_{[4,3,2]}\nonumber\\&&\Theta_{n,1,4,\ell}[q]+ S^*_{[8,1,0]}\Theta_{n,3,12,\ell}[q]+q^{58\ell}S^*_{[7,2,0]}{\rm Tr} (X_1^{[7,2,0]})^n+q^{34\ell}\nonumber\\&&S^*_{[5,3,1]}{\rm Tr} (X_1^{[5,3,1]})^n+q^{48\ell}S^*_{[6,3,0]}{\rm Tr} (X_3^{[6,3,0]})^n.{\label{HW33}}\end{eqnarray}
Where the explicit forms of $X_1^{[7,2,0]}$, $X_2^{[5,3,1]}$, and $X_3^{[6,3,0]}$ are given in Appendix \ref{app0} and algebraic polynomial forms of \ref{HW33} are presented in Appendix~\ref{app1}.
\subsection{$[4]$-colored HOMFLY-PT polynomial of the knots $\hat{Q}_3(1,-1,n,\ell)$\label{t.symrac1}}
In this case, the tensor product decomposition rules are as follows:
\begin{eqnarray*}\bigotimes^3 [4] &=& [12,0,0]\bigoplus [10,1,1]\bigoplus  [8,2,2]\bigoplus  [6,6,0]\bigoplus  [6,3,3]\bigoplus  [5,2,2]\bigoplus [4,4,4]\\&&\bigoplus 2 [11,1,0]\bigoplus
 2[9,2,1]\bigoplus 2 [7,3,2]\bigoplus 2 [6,5,1]\bigoplus 2 [5,4,3]\bigoplus 3 [10,2,0]\\&& \bigoplus 3 [7,5,0]\bigoplus 3 [7,4,1]\bigoplus 3 [6,4,2]\bigoplus 4 [7,4,1]\bigoplus 4 [9,3,0]\bigoplus 5 [8,4,0].\end{eqnarray*}
Thus, we have one $5\times 5$-matrix, two $4\times 4$-matrices, four $3\times 3$-matrices, five $2\times 2$-matrices and  six $1\times 1$-matrices.
{\small \begin{table}[h]
\begin{center}
\begin{tabular}{ |p{3cm}|p{3cm}|p{3cm}|p{3cm}|}
\hline
$\Xi_{\alpha} \in [4]^3$  & Matrix size $(M)$& Trace $\Theta_{n,t,m,\ell}[q]$& Twist factor $\Lambda_{M}$ \\
\hline
[9,2,1] &2  & $\Theta_{n,3,17,\ell}[q]$& ${102 \ell}$\\
\hline
[7,3,2] &2  & $\Theta_{n,2,12,\ell}[q]$& ${72 \ell}$\\
\hline
[11,1,0] &2  & $\Theta_{n,4,24,\ell}[q]$& ${144 \ell}$\\
\hline
[6,5,1]& 2 & $\Theta_{n,2,12,\ell}[q]$& ${72 \ell}$\\
\hline
[5,4,1]& 2 & $\Theta_{n,1,9,\ell}[q]$& ${54 \ell}$\\
\hline
[10,2,0]& 3 & To be determined & ${124\ell}$\\
\hline
[8,3,1]& 3 & To be determined & ${88 \ell}$\\
\hline
[7,5,0]& 3 & To be determined & ${88\ell}$\\
\hline
[6,4,2]& 3 & To be determined & ${64\ell}$\\
\hline
[9,3,0]& 4 & To be determined & ${108 \ell}$\\
\hline
[7,4,1]& 4 & To be determined  & ${78 \ell}$\\
\hline
[8,4,0]& 5 & To be determined & ${96 \ell}$\\
\hline
\end{tabular}
\caption{Twisted trace and twisted factors table for $\Xi_{\alpha} \in  [4]^{\otimes 3}$. }\label{table5}
\end{center}
\end{table}
}\\

The braiding matrices in this case are

\begin{equation*}{\label{col3}}
\hat{\mathfrak{R}}^{[11,1,0]}=\left(\begin{array}{cc}
 -q^{20} & 0 \\
 0 & q^{28} \\
\end{array}
\right),\hat{\mathfrak{R}}^{[9,2,1]}=\left(
\begin{array}{cc}
 q^{14} & 0 \\
 0 & -q^{20} \\
\end{array}
\right),\hat{\mathfrak{R}}^{[5,4,3]}=\left(
\begin{array}{cc}
 q^8 & 0 \\
 0 & -q^{10} \\
\end{array}
\right),\end{equation*}

\begin{equation*}{\label{col31}}
\hat{\mathfrak{R}}^{[7,3,2]}=\hat{\mathfrak{R}}^{[6,5,1]}=\left(
\begin{array}{cc}
 -q^{10} & 0 \\
 0 & q^{14} \\
\end{array}
\right),\hat{\mathfrak{R}}^{[10,2,0]}=\left(
\begin{array}{ccc}
 q^{14} & 0 & 0 \\
 0 & -q^{20} & 0 \\
 0 & 0 & q^{28} \\
\end{array}
\right),
\end{equation*}
\begin{equation*}{\label{col32}}
\hat{\mathfrak{R}}^{[6,4,2]}=\left(
\begin{array}{ccc}
 q^8 & 0 & 0 \\
 0 & -q^{10} & 0 \\
 0 & 0 & q^{14} \\
\end{array}
\right),~\hat{\mathfrak{R}}^{[7,5,0]}=\hat{\mathfrak{R}}^{[8,3,1]}=\left(
\begin{array}{ccc}
 -q^{10} & 0 & 0 \\
 0 & q^{14} & 0 \\
 0 & 0 & -q^{20} \\
\end{array}
\right),
\end{equation*}
\begin{equation*}
\hat{\mathfrak{R}}^{[9,3,0]}=\left(
\begin{array}{cccc}
 -q^{10} & 0 & 0 & 0 \\
 0 & q^{14} & 0 & 0 \\
 0 & 0 & -q^{20} & 0 \\
 0 & 0 & 0 & q^{28} \\
\end{array}
\right),
\end{equation*}
\begin{equation}\hat{\mathfrak{R}}^{[7,4,1]}=\left(
\begin{array}{cccc}
 q^8 & 0 & 0 & 0 \\
 0 & -q^{10} & 0 & 0 \\
 0 & 0 & q^{14} & 0 \\
 0 & 0 & 0 & -q^{20} \\
\end{array}
\right),~\hat{\mathfrak{R}}^{[8,4,0]}=\left(
\begin{array}{ccccc}
 q^8 & 0 & 0 & 0 &0\\
 0 & -q^{10} & 0 & 0&0 \\
 0 & 0 & q^{14} & 0 &0\\
 0 & 0 & 0 & -q^{20}&0 \\
 0 & 0 & 0 & 0&q^{28} \\
\end{array}
\right),~
\end{equation}
and $U^{\Xi_{\alpha}}$ matrices computed from Equation \ref{symrac}.
From Equation  \ref{HOMFLY}, the  $[4]$-colored  HOMFLY-PT polynomial of  $\hat{Q}_3(1,-1,n,\ell)$ can be expressed as follows:
\begin{eqnarray}
\mathcal{H}_{[4]}^{\hat{Q}_3(1,-1,n,\ell)} &=& \frac{A^{-24\ell}q^{-144\ell}}{S^*_{[4}}\sum_{\alpha} {\rm Tr}_{\Xi_{\alpha}}  (\hat{\mathfrak{R}}^{\Xi_{\alpha}}_1)(\hat{\mathfrak{R}}^{\Xi_{\alpha}}_2)^{-1})^n(\hat{\mathfrak{R}}^{\Xi_{\alpha}}_1\hat{\mathfrak{R}}^{\Xi_{\alpha}}_2)^{3\ell}~,\nonumber\\&=& \frac{1}{S^*_{[4]}} (q^{168\ell}S^*_{[12]}+(-q)^{120\ell}S^*_{[10,1,1]}+q^{84\ell}S^*_{[8,2,2]}+q^{84\ell}S^*_{[6,6]}+(-q)^{60\ell}\nonumber\\&&S^*_{[6,3,3]}+(-q)^{60\ell}S^*_{[5,5,2]}+S^*_{[11,2,0]}{\rm Tr}_{[11,2,0]} (\hat{\mathfrak{R}}^{[11,2,0]}_1(\hat{\mathfrak{R}}^{[11,2,0]}_2)^{-1})^n\nonumber\\&&(\hat{\mathfrak{R}}^{[11,2,0]}_1\hat{\mathfrak{R}}^{[11,2,0]}_2)^{3\ell}+S^*_{[9,2,1]}{\rm Tr}_{[9,2,1]} (\hat{\mathfrak{R}}^{[9,2,1]}_1(\hat{\mathfrak{R}}^{[9,2,1]}_2)^{-1})^n(\hat{\mathfrak{R}}^{[9,2,1]}_1\hat{\mathfrak{R}}^{[9,2,1]}_2)^{3\ell}\nonumber\\&&+S^*_{[7,3,2]}{\rm Tr}_{[7,3,2]} (\hat{\mathfrak{R}}^{[7,3,2]}_1(\hat{\mathfrak{R}}^{[7,3,2]}_2)^{-1})^n(\hat{\mathfrak{R}}^{[7,3,2]_1}\hat{\mathfrak{R}}^{[7,3,2]}_2)^{3\ell}+S^*_{[6,5,1]}\nonumber\\&&{\rm Tr}_{[6,5,1]} (\hat{\mathfrak{R}}^{[6,5,1]}_1(\hat{\mathfrak{R}}^{[6,5,1]}_2)^{-1})^n(\hat{\mathfrak{R}}^{[6,5,1]}_1\hat{\mathfrak{R}}^{[6,5,1]}_2)^{3\ell}+S^*_{[5,4,3]}{\rm Tr}_{[5,4,3]} \nonumber\\&&(\hat{\mathfrak{R}}^{[5,4,3]}_1(\hat{\mathfrak{R}}^{[5,4,3]}_2)^{-1})^n(\hat{\mathfrak{R}}^{[5,4,3]}_1\hat{\mathfrak{R}}^{[5,4,3]}_2)^{3\ell}+ S^*_{[10,2,0]}{\rm Tr}_{[10,2,0]}  \nonumber\\&&(\hat{\mathfrak{R}}^{[10,2,0]}_1(\hat{\mathfrak{R}}^{[10,2,0]}_2)^{-1})^n (\hat{\mathfrak{R}}^{[10,2,0]}_1\hat{\mathfrak{R}}^{[10,2,0]}_2)^{3\ell}
+S^*_{[7,5,0]}{\rm Tr}_{[7,5,0]}\nonumber\\&&(\hat{\mathfrak{R}}^{[7,5,0]}_1\hat{\mathfrak{R}}^{[7,5,0]}_2)^{-1})^n(\hat{\mathfrak{R}}^{[7,5,0]}_1\hat{\mathfrak{R}}^{[7,5,0]}_2)^{3\ell}
+S^*_{[6,4,2]}{\rm Tr}_{[6,4,2]} \nonumber\\&& (\hat{\mathfrak{R}}^{[6,4,2]}_1(\hat{\mathfrak{R}}^{[6,4,2]}_2)^{-1})^n(\hat{\mathfrak{R}}^{[6,4,2]}_1\hat{\mathfrak{R}}^{[6,4,2]}_2)^{3\ell}+S^*_{[9,3,0]}{\rm Tr}_{[9,3,0]} \nonumber\\&& (\hat{\mathfrak{R}}^{[9,3,0]}_1(\hat{\mathfrak{R}}^{[9,3,0]}_2)^{-1})^n(\hat{\mathfrak{R}}^{[9,3,0]}_1\hat{\mathfrak{R}}^{[9,3,0]}_2)^{3\ell}+S^*_{[7,4,1]}{\rm Tr}_{[7,4,1]} \nonumber\\&& (\hat{\mathfrak{R}}^{[7,4,1]}_1(\hat{\mathfrak{R}}^{[7,4,1]}_2)^{-1})^n(\hat{\mathfrak{R}}^{[7,4,1]}_1\hat{\mathfrak{R}}^{[7,4,1]}_2)^{3\ell}+S^*_{[8,4,0]}{\rm Tr}_{[8,4,0]} \nonumber\\&&(\hat{\mathfrak{R}}^{[8,4,0]}_1(\hat{\mathfrak{R}}^{[8,4,0]}_2)^{-1})^n(\hat{\mathfrak{R}}^{[8,4,0]}_1\hat{\mathfrak{R}}^{[8,4,0]}_2)^{3\ell}.{\label{HW4}}
\end{eqnarray}
Using Table $\ref{table5}$, we can rewrite   Equation \ref{HW4} into  the following
\begin{eqnarray}
\mathcal{H}_{[4]}^{\hat{Q}_3(1,-1,n,\ell)}&=& \frac{A^{-24\ell}q^{-144\ell}}{S^*_{[4]}} (q^{168\ell}S^*_{[12]}+(-q)^{120\ell}S^*_{[10,1,1]}+q^{84\ell}S^*_{[8,2,2]}+q^{84\ell}S^*_{[6,6]}+\nonumber\\&&(-q)^{60\ell}S^*_{[6,3,3]}+(-q)^{60\ell}S^*_{[5,5,2]}+S^*_{[11,1,0]} \Theta_{n,4,24,\ell}[q]+S^*_{[9,2,1]} \Theta_{n,3,17,\ell}[q]+\nonumber\\&&S^*_{[7,3,2]}) \Theta_{n,2,12,\ell}[q]+S^*_{[6,5,1]} \Theta_{n,2,12,\ell}[q]+
S^*_{[5,4,3]}\Theta_{n,1,9,\ell}[q]+
q^{124\ell}S^*_{[10,2,0]}\nonumber\\&&{\rm Tr} (X_1^{[10,2,0]})^n+q^{88\ell}S^*_{[7,5,0]}{\rm Tr} (X_2^{[7,5,0]})^n+q^{64\ell}S^*_{[6,4,2]}{\rm Tr} (X_3^{[6,4,2]})^n+q^{108\ell}\nonumber\\&&S^*_{[9,3,0]}{\rm Tr} (X_4^{[9,3,0]})^n+q^{78\ell}S^*_{[7,4,1]}{\rm Tr} (X_5^{[7,4,1]})^n+q^{96\ell}S^*_{[8,4,0]}{\rm Tr} (X_6^{[8,4,0]})^n\nonumber\\&&+q^{88\ell}S^*_{[8,3,1]}{\rm Tr} (X_7^{[8,3,1]})^n.{\label{HW44}}\end{eqnarray}
where the explicit forms of matrices $\{X_i^{\Xi_{\alpha}}\}_{i=1,\ldots, 7}$ are given in Appendix \ref{app0} and the colored HOMFLY-PT polynomial  forms (\ref{HW44}) are presented in Appendix~\ref{app1}.\\

Notice that in this section we have computed the  $[r]$-colored HOMFLY-PT polynomials for small values of $r$.
Despite the technical details, our methods are    straightforward  and  it would be interesting to investigate similar  closed form expressions for higher colors. In the following section, we shall investigate   reformulated invariants in the context of topological string duality.
\section{Integrality structures in topological strings }\label{s5.int}
 Gopakumar and Vafa studied  the duality between $SU(N)$  Chern-Simons theory on the three-dimensional sphere  $S^3$ and closed A-model topological string theory on a resolved conifold $\cal{O}$(-1)  + $\cal{O}$ (-1) over $\mathbf P^1$. In particular, a closed string partition function  on the resolved conifold target space  was shown to be the  Chern-Simons free energy $\ln Z[S^3]$.
\begin{equation}
\ln Z[S^3]=-\sum_g \mathcal{F}_{g}(t) g_{s}^{2-2g},
\end{equation}
 where $\mathcal{F}_{g}(t)$ represents the genus $g$ topological string amplitude, $g_s=\frac{2 \pi}{k+N}$ denotes the string coupling constant and $t=\frac{2 \pi i N}{k+N}$ indicates the ${\rm K\ddot{a}hler}$ parameter of $\mathbf P^1$.
 Ooguri and Vafa verified the topological string duality conjecture in the presence of simplest Wilson loop (unknot) observable \cite{OV}. The test of Gopakumar-Vafa duality was performed for the unknot in \cite{OV}. Later, it was extended to  other knots \cite{OV,LMV, LM1, LM2}. This conjecture which is    also known as {\bf LMOV} conjecture can be stated as follows:
\begin{eqnarray}
\left\langle Z(U,V)\right\rangle_{S^3}&=&\sum_{{\bf R}} {\mathcal{H}}^{\star}_{[{\bf R}]}(q,{\bf A})  Tr_{{\bf R}} V=\exp\bigg[\sum_{n=1}^{\infty} \left(\sum\limits_ {{\bf R}}\frac{1}{n} f_{{\bf R}}({\bf A}^n,q^n)Tr_{{\bf R}} V^n\right)\bigg],\nonumber\\
&&{\rm where~}\nonumber\\f_{\bf {\bf R}}(q, {\bf A})&=&\sum_{i,j} \frac{1}{(q-q^{-1})}{\widetilde {\bf N}}_{{\bf R},i,j}{\bf A}^i q^{j}.\label{guv1}
\end{eqnarray}
Here ${\bf R}$ denotes the irreducible representation of $U(N)$, ${\widetilde {\bf N}}_{R, i,j}$, in the reformulated invariants $f_{\bf R}({\bf A},q)$, are integers, where $i$ and $j$ are the charges and spins, respectively. These integers count the number of D2-brane intersecting D4-brane \cite{GV1, GV2}.
Using group theory method for the powers of holonomy $V$, these reformulated invariants can be written in the terms of colored HOMFLY-PT polynomials (\ref{guv1}). For few lower dimensional representations, the explicit forms are as follows:
{\footnotesize
\begin{eqnarray}
f_{[1]}(q,{\bf A})&=&{\mathcal{H}}^{*}_{[1]}(q,{\bf A}),\nonumber \\
f_{[2]}(q,{\bf A})&=&{\mathcal{H}}^{*}_{[2]}(q,{\bf A})-{1\over 2}\Big({ \mathcal{H}}^{*}_{[1]}(q,{\bf A})^2+{\mathcal{H}}^{*}_{[1]}(q^2,{\bf A}^2)\Big),\nonumber\\
f_{[1^2]}(q,{\bf A})&=&{\mathcal{H}}^{*}_{[1^2]}(q,A)-{1\over 2}\Big( {\mathcal{H}}^{*}_{[1]}(q,A)^2-{\mathcal{H}}^{*}_{[1]}(q^2,A^2)\Big),\nonumber\\
\ldots \nonumber
\end{eqnarray}}
where ${\cal H}^{*}_{[r]}(q,{\bf A})$ is the un-normalized $[r]$-colored HOMFLY-PT polynomial. These reformulated invariant can be equivalently written as \cite{LMV}:
\begin{eqnarray}\label{ic}
f^{\mathcal{K}}_{{\bf R}} (q,{ A})=\sum_{m,k\ge 0,s} C_{{\bf R}{\bf S}}\hat {\bf N}^{\mathcal{K}}_{{\bf S},m,k}{ A}^m(q-q^{-1})^{2k-1},
\end{eqnarray}
where $\hat {\bf N}^{\mathcal{K}}_{{\bf S},m,k}$ are  called refined integers and
\begin{eqnarray}
C_{{\bf R}{\bf S}}={1\over q-q^{-1}}\sum_{\Delta}{1\over z_\Delta}\psi_{{\bf R}}(\Delta)\psi_{{\bf S}}(\Delta){\prod_{i=1}^{l(\Delta)}\Big(q^{\xi_i}-q^{-\xi_i}\Big)}\nonumber.
\end{eqnarray}
 Where $\psi_{{\bf R}}(\Delta)$ denotes the characters of symmetric groups  and $z_{\Delta}$ is the standard symmetric factor of the Young diagram discussed in \cite{Fulton,Mironov:2017hde}. Integrality of the coefficients $\hat {\bf N}_{{\bf S},m,k}$ was first  checked for several  knots in \cite{RS, Borhade:2003cu, Zodinmawia:2011oya, Mironov:2017hde,VRR}. Then a general proof was given  in \cite{liu2010proof}.  The refined integers $\hat {\bf N}_{{\bf S},m,k}$ for quasi-alternating knots $\hat{Q}_3(1,-1,n,\pm1)$ are given in the following proposition.  We have:\\
\textbf{Proposition 4.} \emph{ Let $n\geq4$  be an integer.  The refined BPS integers $\hat{\bf N}^{\hat{Q}_3(1,-1,n,\pm
1)}_{[1],\mp5,k}$ and $\hat{\bf N}^{\hat{Q}_3(1,-1,n,\pm1)}_{[1],\mp7,k}$ for quasi-alternating knot $\hat{Q}_3(1,-1,n,\pm1)$ are given by\\
 \begin{eqnarray}{\label{CHP}}
\hat{\bf N}^{\hat{Q}_3(1,-1,n,-1)}_{[ 1],5,k}&=& G_{n,k},~~~~ \hat{\bf N}^{\hat{Q}_3(1,-1,n,-1)}_{[ 1],7,k}= F_{n,k},\nonumber\\~~
\hat{\bf N}^{\hat{Q}_3(1,-1,n,+1)}_{[1],-5,k}&=&
-G_{n,k},~~~~ \hat{\bf N}^{\hat{Q}_3(1,-1,n,+1)}_{[ 1],-7,k}=- F_{n,k},\nonumber\\
\end{eqnarray}}
~\text{where},~
 \begin{eqnarray*}{\label{CHP1}}
F_{n,k}&=& \left(\frac{2 (-1)^{n-1}}{2k!}\prod_{i=0}^{k-1}(n-i)(n+i) \setlength\arraycolsep{1pt}
{}_2{ \bf F}_1\left[\begin{matrix}k-n&, &k+n&, &\frac{2k+1}{2}\end{matrix};\frac{1}{4}\right]-\left(\delta_{k,0}+3\delta_{k,1}+\delta_{k,2}\right)\right), \nonumber
\end{eqnarray*}
 \begin{eqnarray*}{\label{CHP2}}
G_{n,k}&=& \left(\frac{2(-1)^{n}}{2k!}\prod_{i=0}^{k-1}(n-i)(n+i)  \setlength\arraycolsep{1pt}
{}_2{ \bf F}_1\left[\begin{matrix}k-n&, &k+n&, &\frac{2k+1}{2}\end{matrix};\frac{1}{4}\right]+\left(\delta_{k,0}+6\delta_{k,1}+5\delta_{k,2}+\delta_{k,3}\right) \right),
\end{eqnarray*}
here $ \setlength\arraycolsep{1pt}
{}_2{ \bf F}_1\left[\begin{matrix}a&, &b&, &c\end{matrix};z\right]$  represents the hyper-geometric function and $k$ takes values from $1$ to $n$. For clarity, we shall list  $F_{n,k}$ and $G_{n,k}$ for few values of $n$ and $k\in\{1,2,\ldots, n\}$:
\begin{eqnarray*}
F_{4,k}&=&\{1,-3,-4,-1\},\\
G_{4,k}&=&\{2, 7 ,5, 1\},\\
G_{7,k}&=&\{-1, 12, 22, 0, -14, -7, -1\},\\
F_{7,k}&=&\{4, -8, -21, 0, 14, 7, 1\},\\
G_{13,k}&=&\{-7, 31, 131, -65, -377, -78, 390, 273, -65, -143, -65, -13, -1\},\\
F_{13,k}&=&\{10, -27, -130, 65, 377, 78, -390, -273, 65, 143, 65, 13, 1\}.
\end{eqnarray*}
To the best of our knowledge, finding a closed formula  for refined integers of other charges and $[r]>1$ is still an open problem.  Indeed, we checked other properties of $\hat{\bf N}$ up to the level $|{\bf S}|=2$ for the knot  $\hat{Q}_3(1,-1,n,\pm 1 )$. They satisfy the following
\begin{eqnarray*}
\sum_m \hat{\bf N}^{\hat{Q}_3(1,-1,n,\pm 1)}_{{\bf S},m,k}&=0.&
\end{eqnarray*}
Note that, throughout the rest of this paper, we find it more convenient  to use the  notation   $\hat{Q}_3(n,\pm1)$ instead of $\hat{Q}_3(1,-1,n,\pm1)$.
We present refined integers for the  knots $ \hat{Q}_3(2,-1),\;\hat{Q}_3(4,-1)={\bf 10_{157}},\; \hat{Q}_3(5,-1 )$ and $\hat{Q}_3(10,-1)$, with $[r]=[1]$.\\
\begin{center}
\begin{tabular}{cccc}
$\hat{\bf N}^{\hat{Q}_3(2,-1)}_{[ 1]}:$ &
\begin{tabular}{|c|cccc|}
\hline
&&&&\\
$ k \backslash m=$ & 3 & 5 & 7 & 9 \\
&&&&\\
\hline
&&&&\\
0&-4 & 8 & -5 & 1 \\
1& -4 & 6 & -2 & 0 \\
2& -1 & 1 & 0 & 0 \\
&&&&\\
\hline
\end{tabular}
\end{tabular}
\end{center},

\begin{center}
\begin{tabular}{cccc}
$\hat{\bf N}^{\hat{Q}_3(4,-1)}_{[ 1]}:$ &
\begin{tabular}{|c|cccc|}
\hline
&&&&\\
$ k \backslash m=$ & 3 & 5 & 7 & 9 \\
&&&&\\
\hline
&&&&\\
0&-2 & 2 & 1 & -1 \\
1& -5 & 7 & -3 & 1 \\
2& -2 & 5 & -4 & 1 \\
3& 0 & 1 & -1 & 0 \\
&&&&\\
\hline
\end{tabular}, ~
$\hat{\bf N}^{\hat{Q}_3(5,-1)}_{[ 1]}:$ &
\begin{tabular}{|c|cccc|}
\hline
&&&&\\
$ k \backslash m=$ & 3 & 5 & 7 & 9 \\
&&&&\\
\hline
&&&&\\
0&  -5 & 11 & -8 & 2 \\
1& -5 & 10 & -6 & 1 \\
2& 1 & -4 & 5 & -2 \\
3& 1 & -5 & 5 & -1 \\
4& 0 & -1 & 1 & 0 \\&&&&\\
\hline
\end{tabular}
\end{tabular},
\end{center}
\begin{center}\begin{tabular}{cccc}
$\hat{\bf N}^{\hat{Q}_3(10,-1)}_{[ 1]}$ &
\begin{tabular}{|c|cccc|}
\hline
&&&&\\
$ k \backslash m=$ & 3 & 5 & 7 & 9 \\
&&&&\\
\hline
&&&&\\
0& 0 & -4 & 7 & -3 \\
1& -10 & 20 & -16 & 6 \\
2& -19 & 61 & -60 & 18 \\
3& 11 & -15 & 15 & -11 \\
4& 29 & -98 & 98 & -29 \\
5& 2 & -35 & 35 & -2 \\
6& -14 & 40 & -40 & 14 \\
7& -7 & 35 & -35 & 7 \\
8& -1 & 10 & -10 & 1 \\
9& 0 & 1 & -1 & 0 \\
&&&&\\

\hline
\end{tabular}
\end{tabular},~\begin{tabular}{cccc}
$\hat{\bf N}^{\hat{Q}_3(11,-1)}_{[ 1]}$ &
\begin{tabular}{|c|cccc|}
\hline
&&&&\\
$ k \backslash m=$ & 3 & 5 & 7 & 9 \\
&&&&\\
\hline
&&&&\\
 0&-7 & 17 & -14 & 4 \\
 1&-10 & 27 & -23 & 6 \\
 2&23 & -65 & 66 & -24 \\
 3&25 & -99 & 99 & -25 \\
 4&-34 & 77 & -77 & 34 \\
 5&-40 & 154 & -154 & 40 \\
 6&6 & 22 & -22 & -6 \\
 7&20 & -66 & 66 & -20 \\
 8&8 & -44 & 44 & -8 \\
 9&1 & -11 & 11 & -1 \\
 10&0 & -1 & 1 & 0 \\&&&&\\
\hline
\end{tabular}
\end{tabular}.
\end{center}
Finally, it is worth mentioning that refined integers for representations whose length $|{\bf{R}}|= 2$  are presented in Appendix ~\ref{app2}.
\section{Conclusion and discussion}\label{s6.int}
In this paper, we studied the twisted generalized hybrid weaving knots ${\hat Q}_3(m_1,-m_2, n,\ell)$ which are  obtained  by taking  closures of  the 3-braids of type   $(\sigma_1^{m_{1}} \sigma_2^{-m_2})^n \left(\sigma_1 \sigma_2\right)^{3\ell}$. One of the features of  this family  of knots is that it  contains a large class of quasi-alternating knots. We used the modified Reshtikhin-Turaev formalism \cite{RT1}-\cite{RT2} to obtain a closed form algebraic formula for the  $[r]$-colored HOMFLY-PT polynomials of these type of knots. In Proposition 1, we gave  a general formula for the trace term ( \ref{Tracemod}) using  $\hat{\mathfrak{R}_i}$-matrices. Interestingly, for the special case $m_1=m_2=m$, the  sum of the absolute values of the coefficients of  $\Omega((m,m,n,\ell);q)$ in trace term ( \ref{Tracemod1}) is related to generalized Lucas numbers $L_{m,2n}$, see ( \ref{RLucas}). Using the  trace result ( \ref{Tracemod}), we have been able to give an  explicit closed form expression of the HOMFLY-PT polynomial of  twisted generalized hybrid weaving knot $\hat{Q}_3(m_1,-m_2,n,\ell)$ ( \ref{HWK}).
In addition, for the quasi-alternating knots $\hat{Q}_3(1,-1,n,\ell)$, we have explicitly computed the exact coefficients of the Jones and the Alexander polynomials. For quasi-alternating knots, these coefficients  are, respectively,  the ranks of the Khovanov homology and link Floer homology. Moreover, we conjectured that the  determinants of twisted hybrid weaving knots can be expressed in terms of generalized Lucas numbers. Furthermore,
we proved that the asymptotic nature of the coefficients of the Alexander polynomials of quasi-alternating knots satisfy  Fox trapezoidal Conjecture.
Motivated by the Laurent polynomial structure  for HOMFLY-PT polynomial of hybrid weaving knots obtained in  \cite{VRR}, we generalized such structure $\Theta_{n,t,m,\ell}[q]$ ( \ref{conj3}) to the wider class of twisted generalized hybrid weaving knots. In  Section  \ref{s4.int}, we have computed the colored HOMFLY-PT polynomial for the knots  $\hat{Q}_3(1,-1,n,\ell)$ up to $[r]=4$ and tabulated them in Appendix~ \ref{app1}. Our
methods can be straightforwardly extended  to write  the polynomial form for higher colors $[r]>4$. With this polynomial data, we validated the reformulated invariants and found that  certain BPS integers can be expressed in terms of Hyper-geometric functions for $\hat{Q}_{3}(1,-1,n,\ell)$, see ( \ref{CHP}).\\
At this stage, we have investigated the Laurent polynomial   structure $\Theta_{n,t,m,\ell}[q]$ in the  two-dimensional case.  Similar expressions for higher dimensions  are  to be determined. This will help us  writing the  $[r]$-colored HOMFLY-PT polynomial in algebraic compact polynomial form.  Such an expression  can be used to study the  head and tail of the colored Jones polynomial of quasi-alternating links, as well as new quantum invariants (associated with  knot complements) \cite{costantino2014,gukov2020,gukov2017} and  A-polynomials. On the other hand, we plan  to discuss the  generalization  of this study to $m$-strand braids with  $m>3$.

\vspace{0.3cm}

\textbf{Acknowledgements.} This research was funded  by  United Arab Emirates University, UPAR grant $\#$G00003290.\\
The authors  would like to thank P. Ramadevi, R. Mishra, R. Staffeldt, A. Mironov, A. Morozov, A. Sleptsov, and P.Sulkowski for stimulating discussions and correspondence.

\section{The matrices $X_i^{\Xi_{\alpha}}$}{\label{app0}}
In this appendix, we list the  explicit forms of the  matrices $X_i^{\Xi_{\alpha}}$ in Equations   \ref{HW33} and  \ref{HW44}.
\begin{equation*}
{X^{[6,3,0]}=\left(
\begin{array}{cccc}
 -\frac{1}{q^{12} \left(1+q^2+q^4+q^6\right)} & \frac{m_{12}}{q^{11}} & -\frac{m_{13}}{q^6} & \frac{m_{14}}{q^3} \\
 -\frac{m_{12}}{q^9} & \frac{1+2 q^2+2 q^4+q^6+2 q^8+2 q^{10}+q^{12}}{q^8 \left(1+q^2\right) \left(1+q^4\right) \left(1+q^2+q^4+q^6+q^8\right)}
& -\frac{m_{23}}{q^5} & -q^2 m_{24} \\
 -m_{13} & \frac{m_{23}}{q} & \frac{q^4 \left(1+q^2-q^4+q^6+q^8\right)}{1+q^4+q^6+q^{10}} & q^{13} m_{34} \\
 -q^9m_{14}& -q^{12}m_{24} & -q^{19} m_{34} & -m_{44}
\\
\end{array}
\right)},
\end{equation*}

\begin{eqnarray*}
m_{34}&=&\frac{\sqrt{1+q^2 \left(1+q^2\right) \left(2+q^2+3 q^4+2 q^6+3 q^8+2 q^{10}+2 q^{12}+q^{14}+q^{16}\right)}}{\left(1+q^2\right)
\left(1+q^4\right) \left(1-q^2+q^4\right) \left(1+q^2+q^4+q^6+q^8\right)},\\
m_{24}&=&\frac{\left(1+q^2+q^4\right) \sqrt{1+q^2 \left(1+q^2\right) \left(2+q^2+2 q^4+q^6+2 q^8+q^{10}+q^{12}\right)}}{\left(1+q^2\right)
\left(1+q^4\right) \left(1+q^2+q^4+q^6+q^8\right)},\\
m_{23}&=&\frac{\left(1-q^4+q^8\right) \sqrt{1+q^2 \left(1+q^2\right) \left(2+q^2+2 q^4+q^6+q^8\right)}}{\left(1+q^2\right)
\left(1+q^4\right) \left(1+q^2+q^4+q^6+q^8\right)},\\
m_{12}&=&\frac{\sqrt{1+q^2+q^4}}{ \left(1+q^2+q^4+q^6\right)},\\
m_{13}&=&\frac{\sqrt{1+q^2+q^4+q^6+q^8}}{q^2+q^4+q^6+q^8},\\
m_{44}&=&\frac{q^{30}}{\left(1+q^2\right) \left(1+q^4\right) \left(1-q^2+q^4\right) \left(1+q^2+q^4+q^6+q^8\right)}.
\end{eqnarray*}

\begin{equation*}
X^{[5,3,1]}=\left(
\begin{array}{ccc}
 \frac{1}{q^6+q^8+q^{10}} & -\frac{1}{q^5 \sqrt{1+q^2+q^4}} & \frac{\sqrt{1+q^2+q^4+q^6+q^8}}{q^2+q^4+q^6} \\
 \frac{1}{q^3 \sqrt{1+q^2+q^4}} & -\frac{1}{q^2}+\frac{1}{1+q^4} & -\frac{q^3 \sqrt{\frac{1+q^2+q^4+q^6+q^8}{1+q^2+q^4}}}{1+q^4} \\
 \frac{q^4 \sqrt{1+q^2+q^4+q^6+q^8}}{1+q^2+q^4} & \frac{q^7 \sqrt{\frac{1+q^2+q^4+q^6+q^8}{1+q^2+q^4}}}{1+q^4} & \frac{q^{14}}{1+q^2+2 q^4+q^6+q^8}
\\
\end{array}
\right).
\end{equation*}

\begin{equation*}
X^{[7,2,0]}=\left(
\begin{array}{ccc}
 \frac{q^{22}}{1+q^4+q^6+q^8+q^{12}} & q^{12}n_{12} & -q^6 n_{13} \\
 -q^6 n_{12}& \frac{-1+q^2-q^4+q^6-q^8}{q^2 \left(1+q^4\right) \left(1-q^2+q^4\right)} & -\frac{n_{23}}{q^4 } \\
 -\frac{n_{13}}{q^4} & \frac{n_{23}}{q^8 } & \frac{1+q^2+q^4}{q^{10} \left(1+q^4\right) \left(1+q^2+q^4+q^6+q^8\right)} \\
\end{array}
\right).
\end{equation*}

\begin{eqnarray*}
n_{12}&=&\frac{ \sqrt{\frac{1+q^8}{1+q^2+q^4+q^6+q^8}}}{1-q^2+q^4},\\
n_{13}&=&\frac{\sqrt{\frac{1+q^2+q^4+q^6+2 q^8+2 q^{10}+2 q^{12}+q^{14}+q^{16}+q^{18}+q^{20}}{1-q^2+q^4}}}{1+q^2+q^4+q^6+q^8},\\
n_{23}&=&\frac{\sqrt{\left(1+q^4+q^6+q^8+q^{12}\right) \left(1+q^2+q^4+q^6+q^8+q^{10}+q^{12}\right)}}{\left(1+2 q^4+q^6+2
q^8+q^{10}+2 q^{12}+q^{16}\right)}.
\end{eqnarray*}

\begin{equation*}
X^{[7,5,0]}=\left(
\begin{array}{ccc}
 \frac{1+q^2+q^4}{q^{10} \left(1+q^4\right) \left(1+q^2+q^4+q^6+q^8\right)} & -x_{11}& \frac{\left(1+q^2+q^4\right) \sqrt{\frac{1+q^8}{1+q^4+q^8}}}{q^2+q^4+q^6+q^8+q^{10}}
\\
 \frac{\sqrt{\frac{1+q^2+q^4+q^6+q^8+q^{10}+q^{12}}{1+q^4+q^8}}}{q^5+q^9} & \frac{-1+q^2-q^4+q^6-q^8}{q^2 \left(1+q^4\right) \left(1-q^2+q^4\right)}
& -x_{23} \\
 x_{31} & \frac{q^{11} \sqrt{\left(1+q^8\right) \left(1+q^2+q^4+q^6+q^8+q^{10}+q^{12}\right)}}{1+q^4+q^6+q^8+q^{12}} & \frac{q^{22}}{1+q^4+q^6+q^8+q^{12}}
\\
\end{array}
\right),
\end{equation*}

\begin{eqnarray*}
x_{11}&=&\frac{\left(1+q^2+q^4+q^6+q^8+q^{10}+q^{12}\right) \sqrt{\frac{1+q^2+q^4}{1+q^4+q^6+q^8+q^{10}+q^{12}+q^{16}}}}{q^7
\left(1+q^4\right) \left(1+q^2+q^4+q^6+q^8\right)},\\
x_{23}&=&\frac{q^7 \sqrt{\left(1+q^8\right) \left(1+q^2+q^4+q^6+q^8+q^{10}+q^{12}\right)}}{1+q^4+q^6+q^8+q^{10}+q^{12}+q^{16}},\\
x_{31}&=&\frac{q^4 \sqrt{\left(1+q^8\right) \left(1+q^4+q^8\right)} \left(1+q^2+q^4+q^6+q^8+q^{10}+q^{12}\right)}{\left(1-q^2+q^4\right)
\left(1+q^2+q^4\right) \left(1+q^2+q^4+q^6+q^8\right)}.
\end{eqnarray*}

\begin{equation}
X^{[10,2,0]}=\left(
\begin{array}{ccc}
 \frac{1+q^4}{q^{14} \left(1+q^4+q^6+q^8+q^{10}+q^{12}+q^{16}\right)} & -y_{12}&y_{13}\\
 \frac{\sqrt{\frac{1-q^2+q^4-q^6+q^8}{1+q^8}}}{q^6-q^8+q^{10}} & y_{22} & -y_{23} \\
 y_{31} &y_{32} & \frac{q^{30} \left(1+q^2+q^4\right)}{\left(1+q^8\right) \left(1+q^2+q^4+q^6+q^8+q^{10}+q^{12}\right)} \\
\end{array}
\right),
\end{equation}

\begin{eqnarray*}
y_{12}&=&\frac{\sqrt{\frac{1-q^2+q^4-q^6+q^8}{1+q^8}} \left(1+q^2+q^4+q^6+q^8\right)}{q^{10} \left(1+q^4+q^6+q^8+q^{10}+q^{12}+q^{16}\right)},\\
y_{13}&=&\frac{\left(1+q^2+q^4+q^6+q^8\right) \sqrt{\frac{c
\left(1+q^4+q^8+q^{10}+q^{12}+q^{14}+q^{16}+q^{18}+q^{20}+q^{24}+q^{28}\right)}{1+q^8}}}{q^4 \left(1-q^2+q^4\right) \left(1+q^2+q^4+q^6+q^8+q^{10}+q^{12}\right)
g_1},\\
y_{22}&=&\frac{-1+q^2-q^4+q^6-q^8+q^{10}-q^{12}}{q^2 \left(1-q^2+q^4\right) \left(1+q^8\right)},\\
y_{23}&=&\frac{q^{10} \sqrt{c d}}{\left(1-q^2+q^4\right)
\left(1+q^2+q^4\right) \left(1+q^8\right) \left(1+q^6+q^{12}\right)},\\
y_{31}&=&\frac{q^6 \sqrt{\frac{c \left(1+q^4+q^8+q^{10}+q^{12}+q^{14}+q^{16}+q^{18}+q^{20}+q^{24}+q^{28}\right)}{1+q^8}}}{1+q^4+q^6+q^8+q^{10}+q^{12}+q^{16}},\\
y_{32}&=&\frac{q^{16} \sqrt{c d}}{\left(1+q^8\right)
\left(1+q^4+q^6+q^8+q^{10}+q^{12}+q^{16}\right)},\\
c&=&\left(1+q^4+q^6+q^8+q^{10}+q^{12}+q^{14}+q^{16}+q^{20}\right),\\
d&=&\left(1+q^2+q^4+q^6+q^8+q^{10}+q^{12}+q^{14}+q^{16}+q^{18}+q^{20}\right),\\
g_1&=&\left(1+q^2+q^4+q^6+q^8+q^{10}+q^{12}+q^{14}+q^{16}\right).
\end{eqnarray*}

\begin{equation*}
X^{[6,4,2]}=\left(
\begin{array}{ccc}
 \frac{1}{q^6+q^8+q^{10}} & -\frac{1}{q^4+q^6+q^8} & \frac{1}{1+q^2+q^4} \\
 \frac{1}{q^4} & \frac{-1+q^2-q^4}{q^2+q^6} & -\frac{q^4}{1+q^4} \\
 \frac{q^2+q^4+q^6+q^8+q^{10}}{1+q^2+q^4} & \frac{q^6 \left(1+q^2+q^4+q^6+q^8\right)}{\left(1+q^4\right) \left(1+q^2+q^4\right)} & \frac{q^{14}}{1+q^2+2
q^4+q^6+q^8} \\
\end{array}
\right).
\end{equation*}
{\tiny\begin{equation*}
X^{[8,4,0]}=\left(
\begin{array}{ccccc}
 \frac{k_0}{q^{20} } & -\frac{\text{k0}}{q^{18} } & \frac{k_0}{q^{14} } & -\frac{k_0}{q^8 } & k_0 \\
 \frac{\left(1+q^2+q^4\right)k_0}{q^{18}} & -\frac{1+2 q^4-q^6+2 q^8+q^{12}}{q^{16} k_1} & \frac{1+q^4-q^6+q^8+q^{12}}{q^{12}k_1}
& \frac{-1+q^2+q^6-q^8}{q^6 k_2} & -\frac{q^4+q^8}{k_2} \\
 \frac{1}{q^{12}} & \frac{-1+q^2-q^4+q^6-q^8}{q^{10} \left(1+q^4\right) \left(1-q^2+q^4\right)} & -k_{33}& \frac{q^2+q^4+q^{12}+q^{14}}{k_3}
& \frac{q^{14}+q^{18}}{k_3} \\
 \frac{1+q^2+q^4+q^6+q^8+q^{10}+q^{12}}{q^2+q^4+q^6+q^8+q^{10}} & \frac{-1+q^2+q^{14}-q^{16}}{1+q^4+q^6+q^8+q^{12}} & -\frac{q^6+q^8+q^{16}+q^{18}}{k_2}
& -\frac{q^{16} k_5}{k_4} & -\frac{q^{30}}{k_4} \\
 k_{51} & \frac{q^{16} k_3}{1+q^4+q^6+q^8+q^{12}} & k_{53} &k_{54} &k_{55} \\
\end{array}
\right),
\end{equation*}}

\begin{eqnarray*}
k_0&=&\frac{1}{\left(1+q^2+q^4+q^6+q^8\right)},\\
k_1&=&\left(1+2 q^4+q^6+2 q^8+q^{10}+2 q^{12}+q^{16}\right),\\
k_2&=&\left(1+q^4+q^6+q^8+q^{12}\right),\\
k_3&=&\left(1+q^4+q^6+q^8+q^{10}+q^{12}+q^{16}\right),\\
k_4&=&\left(1+q^4+q^6+2 q^8+2 q^{12}+q^{14}+q^{16}+q^{20}\right),\\
k_5&=&\left(1+q^2+q^4-q^6+q^8+q^{12}\right),\\
k_{33}&=&\frac{-1+q^2+q^6+q^8-q^{10}+q^{12}+q^{14}+q^{18}-q^{20}}{q^6 \left(1+q^4\right) \left(1-q^2+q^4\right) \left(1+q^2+q^4+q^6+q^8+q^{10}+q^{12}\right)},\\
k_{55}&=&\frac{q^{52}}{\left(1-q^2+q^4\right) \left(1+q^8\right) \left(1+q^2+q^4+q^6+q^8\right) \left(1+q^2+q^4+q^6+q^8+q^{10}+q^{12}\right)},\\
k_{54}&=&\frac{q^{36} \left(1+q^2+q^4+q^6+q^8+q^{10}+q^{12}+q^{14}+q^{16}\right)}{\left(1-q^2+q^4\right) \left(1+q^8\right)
\left(1+q^2+q^4+q^6+q^8\right) \left(1+q^2+q^4+q^6+q^8+q^{10}+q^{12}\right)},\\
k_{51}&=&\frac{q^{12} \left(1+q^2+q^4+q^6+q^8+q^{10}+q^{12}+q^{14}+q^{16}\right)}{1+q^2+q^4+q^6+q^8},\\
k_{53}&=&\frac{q^{24} \left(1+q^4\right) \left(1+q^2+q^4+q^6+q^8+q^{10}+q^{12}+q^{14}+q^{16}\right)}{\left(1-q^2+q^4\right)
\left(1+q^2+q^4+q^6+q^8\right) \left(1+q^2+q^4+q^6+q^8+q^{10}+q^{12}\right)}.
\end{eqnarray*}

{\tiny \begin{equation*}
X^{[8,3,1]}=\left(
\begin{array}{ccc}
 \frac{1+q^2+q^4}{q^{10} \left(1+q^4\right) \left(1+q^2+q^4+q^6+q^8\right)} & -\frac{\left(1+q^2+q^4\right) \sqrt{1+q^2+q^4+q^6+q^8+q^{10}+q^{12}}}{q^7
\left(1+q^4\right) \sqrt{1+q^4+q^8} \left(1+q^2+q^4+q^6+q^8\right)} & \frac{\left(1+q^2+q^4\right) \sqrt{1+q^8}}{\sqrt{1+q^4+q^8} \left(q^2+q^4+q^6+q^8+q^{10}\right)}
\\
 \frac{\sqrt{1+q^2+q^4+q^6+q^8+q^{10}+q^{12}}}{\sqrt{1+q^4+q^8} \left(q^5+q^9\right)} & \frac{-1+q^2-q^4+q^6-q^8}{q^2 \left(1+q^4\right) \left(1-q^2+q^4\right)}
& -\frac{q^7 \sqrt{1+q^8}}{\left(1-q^2+q^4\right) \sqrt{1+q^2+q^4+q^6+q^8+q^{10}+q^{12}}} \\
 \frac{q^4 \sqrt{1+q^8} \left(1+q^2+q^4+q^6+q^8+q^{10}+q^{12}\right)}{\sqrt{1+q^4+q^8} \left(1+q^2+q^4+q^6+q^8\right)} & \frac{q^{11} \sqrt{1+q^8}
\sqrt{1+q^2+q^4+q^6+q^8+q^{10}+q^{12}}}{1+q^4+q^6+q^8+q^{12}} & \frac{q^{22}}{1+q^4+q^6+q^8+q^{12}} \\
\end{array}
\right).
\end{equation*}}

\section{Colored HOMFLY-PT polynomials}{\label{app1}}
Here, we  tabulate the colored HOMFLY-PT polynomials for    few examples of quasi-alternating knots $\hat{Q}_3(n,\pm1)$ for colors $[r]=2,3,4$, discussed in Section  \ref{s4.int}, see Equations  \ref{HW2}$, \ref{HW33}$ and  \ref{HW44}. These polynomials are better rewritten in the matrix-form $(q^2,A^2)$. Here is an example that illustrates  how the two-variable polynomial is represented in matrix-form. We have computed the polynomial  expression of $[2]$-colored  HOMFLY-PT of $\hat{Q}_3(2,-1)$, i.e., ${\bf 3_1\#3_1}$ knot
\begin{eqnarray*}
H_{[2]}^{\hat{Q}_3(2,-1)}&=&   \frac{A^8}{q^8} - \frac{2 A^{10}}{q^4} +\frac{
 2 A^8 - 2 A^{10}}{q^2} +2 A^8 + A^{12}+ (-4 A^{10} + 4 A^{12}) q^2 + (3 A^8 \\&&- 6 A^{10} +
    A^{12}) q^4 + (2 A^8 - 2 A^{10} + 2 A^{12} - 2 A^{14}) q^6 + (A^8 -
    4 A^{10} \\&&+ 6 A^{12} - 2 A^{14}) q^8 + (2 A^8 - 6 A^{10} +
    4 A^{12}) q^{10}  + (2 A^8 - 2 A^{10}\\&&+ A^{12} - 2 A^{14} +
    A^{16}) q^{12 }+ (-2 A^{10} + 4 A^{12} - 2 A^{14}) \\&&q^{14} + (A^8 - 2 A^{10} +
    A^{12}) q^{16}.\end{eqnarray*}
This is expressed into  matrix-form $(q^2,A^2)$ as follows:
{\small\begin{equation*}
H_{[2]}^{\hat{Q}_3(2,-3)}=A^{8} q^{-8} \left(
\begin{array}{ccccc}
 1 & 0 & 0 & 0 & 0 \\
 0 & 0 & 0 & 0 & 0 \\
 0 & -2 & 0 & 0 & 0 \\
 2 & -2 & 0 & 0 & 0 \\
 2 & 0 & 1 & 0 & 0 \\
 0 & -4 & 4 & 0 & 0 \\
 3 & -6 & 1 & 0 & 0 \\
 2 & -2 & 2 & -2 & 0 \\
 1 & -4 & 6 & -2 & 0 \\
 2 & -6 & 4 & 0 & 0 \\
 2 & -2 & 1 & -2 & 1 \\
 0 & -2 & 4 & -2 & 0 \\
 1 & -2 & 1 & 0 & 0 \\
\end{array}
\right).
\end{equation*}}
Similarly, polynomials of  other quasi-alternating knots are  listed in the matrix form $(q^2,A^2)$:
\begin{eqnarray*}
H_{[2]}^{\hat{Q}_3(4,-1)}&=&A^{8} q^{-12} \left(
\begin{array}{ccccc}
 0 & -1 & 0 & 0 & 0 \\
 1 & -2 & 1 & 0 & 0 \\
 3 & 5 & -2 & 0 & 0 \\
 -6 & 4 & 3 & -1 & 0 \\
 -2 & -17 & 6 & 2 & 0 \\
 14 & 2 & -18 & 2 & 0 \\
 -7 & 28 & 2 & -11 & 1 \\
 -12 & -21 & 32 & 4 & -3 \\
 17 & -20 & -28 & 20 & 0 \\
 -1 & 35 & -22 & -21 & 9 \\
 -12 & -5 & 44 & -13 & -8 \\
 10 & -24 & -11 & 32 & -7 \\
 1 & 20 & -29 & -8 & 15 \\
 -5 & 2 & 26 & -20 & -3 \\
 3 & -11 & 0 & 17 & -9 \\
 0 & 5 & -15 & 2 & 8 \\
 0 & 1 & 8 & -9 & 0 \\
 0 & -1 & 1 & 3 & -3 \\
 0 & 0 & -3 & 2 & 1 \\
 0 & 0 & 1 & -1 & 0 \\
\end{array}
\right),
\end{eqnarray*}
\begin{eqnarray*}
H_{[2]}^{\hat{Q}_3(5,-1)}&=&A^{8} q^{-16}\left(\begin{array}{ccccc}
 0 & 0 & 1 & 0 & 0 \\
 0 & 4 & -4 & 0 & 0 \\
 -5 & -2 & 2 & -1 & 0 \\
 6 & -17 & 8 & 3 & 0 \\
 16 & 22 & -24 & 2 & 0 \\
 -29 & 25 & 17 & -14 & 1 \\
 -4 & -76 & 45 & 14 & -4 \\
 57 & 3 & -84 & 22 & 2 \\
 -34 & 106 & 5 & -60 & 12 \\
 -37 & -95 & 146 & 8 & -22 \\
 72 & -75 & -111 & 91 & -2 \\
 -13 & 134 & -75 & -91 & 45 \\
 -43 & -41 & 193 & -57 & -36 \\
 47 & -101 & -49 & 132 & -29 \\
 1 & 84 & -113 & -44 & 66 \\
 -23 & 1 & 130 & -89 & -19 \\
 16 & -58 & 0 & 83 & -40 \\
 1 & 27 & -74 & 5 & 41 \\
 -4 & 9 & 49 & -53 & -1 \\
 1 & -14 & 10 & 24 & -21 \\
 0 & 2 & -26 & 12 & 12 \\
 0 & 3 & 10 & -15 & 2 \\
 0 & -1 & 3 & 2 & -4 \\
 0 & 0 & -4 & 3 & 1 \\
 0 & 0 & 1 & -1 & 0 \\
\end{array}
\right),
\end{eqnarray*}

\begin{equation*}
H_{[3]}^{\hat{Q}_3(4,-1)}=A^{12} q^{-18} \left(
\begin{array}{ccccccc}
 0 & -2 & 1 & 0 & 0 & 0 & 0 \\
 2 & -3 & 1 & 0 & 0 & 0 & 0 \\
 2 & -1 & -1 & 0 & 0 & 0 & 0 \\
 1 & 13 & -6 & -2 & 0 & 0 & 0 \\
 -12 & 10 & 1 & 1 & 0 & 0 & 0 \\
 -4 & -14 & 19 & -1 & 0 & 0 & 0 \\
 15 & -43 & 14 & 2 & 1 & 0 & 0 \\
 23 & 1 & -30 & 7 & -1 & 0 & 0 \\
 -14 & 68 & -59 & 5 & 0 & 0 & 0 \\
 -40 & 50 & 20 & -21 & 4 & 0 & 0 \\
 1 & -74 & 110 & -38 & 2 & -1 & 0 \\
 50 & -108 & 45 & 25 & -14 & 2 & 0 \\
 19 & 35 & -139 & 94 & -21 & 1 & 0 \\
 -46 & 147 & -142 & 21 & 24 & -4 & 0 \\
 -35 & 24 & 103 & -142 & 59 & -9 & 0 \\
 39 & -142 & 219 & -116 & -4 & 9 & 1 \\
 37 & -80 & -14 & 137 & -105 & 28 & -3 \\
 -22 & 110 & -237 & 213 & -62 & -2 & 0 \\
 -35 & 99 & -81 & -61 & 122 & -51 & 6 \\
 14 & -63 & 197 & -261 & 141 & -32 & 4 \\
 27 & -95 & 127 & -41 & -74 & 67 & -11 \\
 -8 & 34 & -125 & 239 & -198 & 75 & -17 \\
 -17 & 71 & -135 & 109 & -2 & -43 & 17 \\
 5 & -13 & 65 & -166 & 195 & -114 & 28 \\
 9 & -47 & 105 & -133 & 70 & 3 & -7 \\
 0 & 6 & -32 & 96 & -140 & 112 & -42 \\
 -7 & 26 & -70 & 115 & -98 & 40 & -6 \\
 4 & 1 & 10 & -50 & 77 & -84 & 42 \\
 0 & -16 & 39 & -77 & 93 & -56 & 17 \\
 0 & 4 & 0 & 20 & -35 & 42 & -31 \\
 0 & 2 & -24 & 48 & -59 & 54 & -21 \\
 0 & 1 & 2 & -6 & 6 & -19 & 16 \\
 0 & -1 & 6 & -30 & 36 & -31 & 20 \\
 0 & 0 & 1 & 5 & 5 & 0 & -11 \\
 0 & 0 & -1 & 12 & -20 & 18 & -9 \\
 0 & 0 & -2 & -1 & -3 & 3 & 3 \\
 0 & 0 & 1 & -4 & 7 & -9 & 5 \\
 0 & 0 & 0 & -1 & 3 & -2 & 0 \\
 0 & 0 & 0 & 3 & -2 & 2 & -3 \\
 0 & 0 & 0 & -1 & -2 & 2 & 1 \\
 0 & 0 & 0 & 0 & 1 & -1 & 0 \\
\end{array}
\right),~
\end{equation*}
{\tiny \begin{eqnarray*}
H_{[3]}^{\hat{Q}_3(5,-1)}=A^{12} q^{-28} \left(
\begin{array}{ccccccc}
 0 & -1 & 1 & 0 & 0 & 0 & 0 \\
 1 & -1 & 0 & 0 & 0 & 0 & 0 \\
 0 & 0 & 0 & 1 & 0 & 0 & 0 \\
 0 & 11 & -6 & -5 & 0 & 0 & 0 \\
 -11 & 15 & -5 & 1 & 0 & 0 & 0 \\
 -10 & -15 & 13 & 7 & -1 & 0 & 0 \\
 25 & -59 & 26 & 5 & 3 & 0 & 0 \\
 45 & -21 & -18 & -8 & 2 & 0 & 0 \\
 -11 & 135 & -100 & -3 & -5 & 0 & 0 \\
 -121 & 137 & -26 & 16 & -7 & 1 & 0 \\
 -32 & -145 & 204 & -32 & 8 & -3 & 0 \\
 172 & -382 & 201 & -28 & 13 & -1 & 0 \\
 167 & 20 & -260 & 103 & -35 & 5 & 0 \\
 -182 & 573 & -532 & 160 & -27 & 8 & 0 \\
 -294 & 313 & 133 & -189 & 82 & -15 & -1 \\
 106 & -676 & 929 & -445 & 102 & -20 & 4 \\
 419 & -743 & 346 & 93 & -153 & 40 & -2 \\
 25 & 425 & -1073 & 847 & -298 & 57 & -8 \\
 -416 & 1048 & -959 & 239 & 148 & -62 & 2 \\
 -155 & -112 & 940 & -1150 & 621 & -165 & 21 \\
 398 & -1155 & 1591 & -971 & 86 & 63 & 4 \\
 230 & -377 & -298 & 1067 & -887 & 323 & -58 \\
 -267 & 960 & -1731 & 1549 & -560 & 63 & -14 \\
 -242 & 538 & -248 & -622 & 976 & -506 & 98 \\
 203 & -708 & 1649 & -2049 & 1163 & -330 & 72 \\
 210 & -693 & 826 & -125 & -609 & 536 & -145 \\
 -101 & 392 & -1110 & 1846 & -1533 & 674 & -167 \\
 -150 & 523 & -890 & 652 & 111 & -396 & 150 \\
 69 & -210 & 731 & -1542 & 1598 & -933 & 287 \\
 95 & -438 & 892 & -988 & 483 & 47 & -91 \\
 -20 & 69 & -315 & 904 & -1238 & 966 & -366 \\
 -59 & 247 & -614 & 939 & -746 & 255 & -22 \\
 21 & -8 & 130 & -556 & 798 & -771 & 386 \\
 18 & -163 & 436 & -745 & 815 & -477 & 116 \\
 -2 & -2 & 0 & 207 & -382 & 483 & -304 \\
 -7 & 60 & -253 & 524 & -622 & 478 & -180 \\
 -2 & 20 & -25 & -87 & 108 & -221 & 207 \\
 4 & -25 & 124 & -330 & 433 & -380 & 174 \\
 -1 & -14 & 39 & 12 & 21 & 57 & -114 \\
 0 & 9 & -50 & 190 & -258 & 239 & -130 \\
 0 & 6 & -37 & 7 & -52 & 24 & 52 \\
 0 & -1 & 23 & -91 & 133 & -143 & 79 \\
 0 & -3 & 17 & -20 & 49 & -30 & -13 \\
 0 & 1 & -7 & 49 & -58 & 65 & -50 \\
 0 & 0 & -7 & 8 & -37 & 29 & 7 \\
 0 & 0 & 1 & -18 & 26 & -29 & 20 \\
 0 & 0 & 3 & -4 & 16 & -16 & 1 \\
 0 & 0 & -1 & 6 & -7 & 10 & -8 \\
 0 & 0 & 0 & 3 & -7 & 6 & -2 \\
 0 & 0 & 0 & -4 & 1 & -1 & 4 \\
 0 & 0 & 0 & 1 & 3 & -3 & -1 \\
 0 & 0 & 0 & 0 & -1 & 1 & 0 \\
\end{array}
\right),
\end{eqnarray*}}

{ \begin{equation*}
H_{[4]}^{\hat{Q}_3(2,-1)}=A^{16} q^{-16}\left(\begin{array}{ccccccccc}
 1 & 0 & 0 & 0 & 0 & 0 & 0 & 0 & 0 \\
 0 & 0 & 0 & 0 & 0 & 0 & 0 & 0 & 0 \\
 0 & 0 & 0 & 0 & 0 & 0 & 0 & 0 & 0 \\
 0 & 0 & 0 & 0 & 0 & 0 & 0 & 0 & 0 \\
 0 & -2 & 0 & 0 & 0 & 0 & 0 & 0 & 0 \\
 2 & -2 & 0 & 0 & 0 & 0 & 0 & 0 & 0 \\
 2 & -2 & 0 & 0 & 0 & 0 & 0 & 0 & 0 \\
 2 & -2 & 0 & 0 & 0 & 0 & 0 & 0 & 0 \\
 2 & 0 & 1 & 0 & 0 & 0 & 0 & 0 & 0 \\
 0 & -4 & 4 & 0 & 0 & 0 & 0 & 0 & 0 \\
 3 & -8 & 5 & 0 & 0 & 0 & 0 & 0 & 0 \\
 4 & -12 & 8 & 0 & 0 & 0 & 0 & 0 & 0 \\
 7 & -14 & 5 & 0 & 0 & 0 & 0 & 0 & 0 \\
 6 & -10 & 6 & -2 & 0 & 0 & 0 & 0 & 0 \\
 5 & -12 & 11 & -4 & 0 & 0 & 0 & 0 & 0 \\
 6 & -16 & 20 & -10 & 0 & 0 & 0 & 0 & 0 \\
 7 & -26 & 32 & -12 & 0 & 0 & 0 & 0 & 0 \\
 10 & -34 & 36 & -12 & 0 & 0 & 0 & 0 & 0 \\
 12 & -34 & 35 & -14 & 1 & 0 & 0 & 0 & 0 \\
 10 & -36 & 38 & -16 & 4 & 0 & 0 & 0 & 0 \\
 13 & -34 & 44 & -32 & 9 & 0 & 0 & 0 & 0 \\
 10 & -42 & 66 & -46 & 12 & 0 & 0 & 0 & 0 \\
 13 & -52 & 77 & -56 & 18 & 0 & 0 & 0 & 0 \\
 14 & -54 & 88 & -62 & 14 & 0 & 0 & 0 & 0 \\
 14 & -58 & 85 & -58 & 19 & -2 & 0 & 0 & 0 \\
 14 & -54 & 86 & -70 & 28 & -4 & 0 & 0 & 0 \\
 13 & -54 & 96 & -86 & 41 & -10 & 0 & 0 & 0 \\
 12 & -58 & 106 & -104 & 56 & -12 & 0 & 0 & 0 \\
 13 & -56 & 112 & -116 & 59 & -12 & 0 & 0 & 0 \\
 10 & -56 & 112 & -110 & 58 & -14 & 0 & 0 & 0 \\
 11 & -50 & 100 & -110 & 62 & -14 & 1 & 0 & 0 \\
 8 & -46 & 100 & -108 & 66 & -24 & 4 & 0 & 0 \\
 9 & -42 & 91 & -114 & 83 & -32 & 5 & 0 & 0 \\
 6 & -36 & 90 & -116 & 84 & -36 & 8 & 0 & 0 \\
 5 & -32 & 77 & -104 & 85 & -36 & 5 & 0 & 0 \\
 4 & -24 & 66 & -94 & 72 & -30 & 6 & 0 & 0 \\
 3 & -20 & 54 & -78 & 66 & -32 & 7 & 0 & 0 \\
 2 & -16 & 44 & -68 & 64 & -34 & 10 & -2 & 0 \\
 2 & -10 & 34 & -60 & 58 & -36 & 14 & -2 & 0 \\
 0 & -8 & 26 & -44 & 50 & -34 & 12 & -2 & 0 \\
 1 & -4 & 15 & -34 & 39 & -24 & 9 & -2 & 0 \\
 0 & -2 & 12 & -22 & 24 & -20 & 8 & 0 & 0 \\
 0 & -2 & 5 & -14 & 22 & -14 & 5 & -2 & 0 \\
 0 & 0 & 4 & -10 & 12 & -12 & 8 & -2 & 0 \\
 0 & 0 & 1 & -4 & 9 & -10 & 5 & -2 & 1 \\
 0 & 0 & 0 & -2 & 4 & -4 & 4 & -2 & 0 \\
 0 & 0 & 0 & 0 & 1 & -2 & 1 & 0 & 0 \\
\end{array}
\right),
\end{equation*}

{\tiny \begin{equation*}
H_{[4]}^{\hat{Q}_3(4,-1)}=A^{16} q^{-26} \left(
\begin{array}{ccccccccc}
 0 & -1 & 1 & 0 & 0 & 0 & 0 & 0 & 0 \\
 1 & -4 & 2 & 0 & 0 & 0 & 0 & 0 & 0 \\
 3 & -6 & 4 & -1 & 0 & 0 & 0 & 0 & 0 \\
 4 & -1 & -2 & -1 & 0 & 0 & 0 & 0 & 0 \\
 -3 & 11 & -9 & 1 & 0 & 0 & 0 & 0 & 0 \\
 -7 & 30 & -18 & 1 & 0 & 0 & 0 & 0 & 0 \\
 -20 & 19 & -1 & 1 & 1 & 0 & 0 & 0 & 0 \\
 -2 & -33 & 39 & -7 & 3 & 0 & 0 & 0 & 0 \\
 33 & -76 & 63 & -17 & -3 & 0 & 0 & 0 & 0 \\
 36 & -82 & 32 & 6 & -3 & 0 & 0 & 0 & 0 \\
 25 & 47 & -102 & 34 & -4 & 0 & 0 & 0 & 0 \\
 -65 & 186 & -176 & 60 & -3 & -2 & 0 & 0 & 0 \\
 -87 & 163 & -108 & 11 & 21 & 0 & 0 & 0 & 0 \\
 -18 & -21 & 167 & -135 & 19 & 1 & 0 & 0 & 0 \\
 73 & -338 & 420 & -170 & 16 & -1 & 0 & 0 & 0 \\
 163 & -322 & 237 & -59 & -28 & 9 & 0 & 0 & 0 \\
 18 & 40 & -226 & 274 & -111 & 4 & 1 & 0 & 0 \\
 -126 & 434 & -726 & 472 & -67 & 3 & -1 & 0 & 0 \\
 -166 & 532 & -514 & 129 & 33 & -15 & 1 & 0 & 0 \\
 -51 & -56 & 357 & -443 & 251 & -59 & 1 & 0 & 0 \\
 200 & -581 & 990 & -894 & 303 & -19 & 1 & 0 & 0 \\
 158 & -591 & 840 & -373 & -68 & 42 & -2 & 0 & 0 \\
 7 & -5 & -429 & 753 & -463 & 161 & -23 & -1 & 0 \\
 -176 & 737 & -1321 & 1287 & -678 & 156 & -7 & 2 & 0 \\
 -180 & 571 & -940 & 712 & -58 & -124 & 18 & 1 & 0 \\
 83 & -73 & 382 & -998 & 883 & -342 & 69 & -5 & 0 \\
 146 & -680 & 1567 & -1769 & 1057 & -390 & 71 & -2 & 0 \\
 105 & -571 & 927 & -792 & 292 & 118 & -75 & -4 & 0 \\
 -73 & 246 & -516 & 1037 & -1266 & 721 & -170 & 21 & 0 \\
 -155 & 567 & -1437 & 2128 & -1590 & 645 & -186 & 27 & 1 \\
 6 & 368 & -888 & 761 & -284 & -38 & 94 & -16 & -3 \\
 60 & -243 & 709 & -1200 & 1402 & -1096 & 415 & -47 & 0 \\
 85 & -511 & 1177 & -1944 & 2015 & -1057 & 315 & -86 & 6 \\
 -13 & -84 & 530 & -728 & 227 & 121 & -77 & 23 & 1 \\
 -80 & 215 & -627 & 1355 & -1615 & 1274 & -675 & 152 & 1 \\
 0 & 302 & -965 & 1556 & -1847 & 1416 & -574 & 133 & -21 \\
 13 & 0 & -92 & 346 & -258 & -183 & 181 & 3 & -10 \\
 40 & -233 & 497 & -1103 & 1736 & -1479 & 808 & -289 & 23 \\
 -2 & -74 & 547 & -1215 & 1459 & -1315 & 824 & -255 & 31 \\
 -33 & 38 & -49 & 114 & -33 & 115 & -242 & 56 & 34 \\
 6 & 130 & -441 & 794 & -1354 & 1555 & -980 & 347 & -57 \\
 2 & 21 & -151 & 651 & -1135 & 1048 & -762 & 404 & -78 \\
 11 & -84 & 88 & -189 & 393 & -262 & 166 & -108 & -15 \\
 -1 & -16 & 233 & -617 & 928 & -1196 & 1051 & -447 & 65 \\
 -9 & 5 & 33 & -149 & 580 & -840 & 609 & -369 & 140 \\
 5 & 36 & -140 & 169 & -338 & 445 & -237 & 61 & -1 \\
 0 & 4 & -44 & 305 & -680 & 804 & -791 & 515 & -113 \\
 0 & -22 & 4 & 23 & -79 & 436 & -525 & 286 & -123 \\
 0 & 4 & 65 & -195 & 237 & -323 & 332 & -103 & -17 \\
 0 & 0 & 9 & -52 & 324 & -578 & 514 & -377 & 160 \\
 0 & 2 & -32 & 11 & -11 & -79 & 280 & -263 & 92 \\
 0 & 1 & 0 & 89 & -231 & 198 & -208 & 169 & -18 \\
 0 & -1 & -5 & 15 & -39 & 285 & -376 & 235 & -114 \\
 0 & 0 & 7 & -47 & 29 & 19 & -53 & 135 & -90 \\
 0 & 0 & 2 & -2 & 106 & -182 & 105 & -92 & 63 \\
 0 & 0 & -1 & -10 & 6 & -67 & 185 & -180 & 67 \\
 0 & 0 & 0 & 14 & -67 & 0 & 28 & -13 & 38 \\
 0 & 0 & -2 & 7 & 2 & 90 & -102 & 41 & -36 \\
 0 & 0 & 1 & -5 & -2 & 29 & -45 & 79 & -57 \\
 0 & 0 & 0 & 1 & 24 & -39 & -15 & 16 & 13 \\
 0 & 0 & 0 & -5 & 3 & -15 & 51 & -48 & 14 \\
 0 & 0 & 0 & 2 & -15 & -13 & 20 & -13 & 19 \\
 0 & 0 & 0 & 2 & 4 & 18 & -14 & -6 & -4 \\
 0 & 0 & 0 & -1 & -3 & 9 & -8 & 19 & -16 \\
 0 & 0 & 0 & 0 & 4 & -6 & -11 & 10 & 3 \\
 0 & 0 & 0 & 0 & 1 & -1 & 9 & -9 & 0 \\
 0 & 0 & 0 & 0 & -3 & -4 & 4 & -2 & 5 \\
 0 & 0 & 0 & 0 & 1 & 2 & 0 & -3 & 0 \\
 0 & 0 & 0 & 0 & 0 & 2 & -1 & 2 & -3 \\
 0 & 0 & 0 & 0 & 0 & -1 & -2 & 2 & 1 \\
 0 & 0 & 0 & 0 & 0 & 0 & 1 & -1 & 0 \\
\end{array}
\right).
\end{equation*}}

\section{Reformulated integers}{\label{app2}}
In this appendix, we present  the table of refined integers for representations whose length $|{\bf{R}}|= 2$ for certain quasi-alternating knots.
\begin{equation*}
 \hat{\bf N}^{\hat{Q}_3(4,-1)}_{[ 2]}=\left(
\begin{array}{c|ccccccc}
\hline
 k/m&6 & 8 & 10 & 12 & 14 & 16 & 18 \\
\hline
 0&18 & -37 & -23 & 102 & -68 & -1 & 9 \\
 1&98 & -251 & -66 & 681 & -669 & 226 & -19 \\
 2&207 & -752 & 344 & 1411 & -2120 & 1149 & -239 \\
 3&217 & -1305 & 2087 & -368 & -1765 & 1509 & -375 \\
 4&119 & -1435 & 4763 & -6898 & 5069 & -2035 & 417 \\
 5&31 & -1012 & 6106 & -14214 & 15867 & -8642 & 1864 \\
 6&3 & -449 & 4876 & -15493 & 20542 & -11882 & 2403 \\
 7&0 & -119 & 2512 & -10458 & 15492 & -9063 & 1636 \\
 8&0 & -17 & 834 & -4570 & 7356 & -4253 & 650 \\
 9&0 & -1 & 172 & -1294 & 2227 & -1255 & 151 \\
 10&0 & 0 & 20 & -229 & 417 & -227 & 19 \\
 11&0 & 0 & 1 & -23 & 44 & -23 & 1 \\
 12&0 & 0 & 0 & -1 & 2 & -1 & 0 \\
\end{array}
\right),
\end{equation*}

\begin{equation*}
\hat{\bf N}^{\hat{Q}_3(4,-1)}_{[ 1,1]}=\left(
\begin{array}{c|ccccccc}
\hline
 k/m&6& 8 & 10 & 12 & 14 & 16 & 18 \\
\hline
0& -15 & 35 & 4 & -66 & 49 & -1 & -6 \\
1& -69 & 200 & -33 & -387 & 440 & -173 & 22 \\
2& -112 & 468 & -377 & -565 & 1095 & -655 & 146 \\
3& -80 & 618 & -1288 & 860 & 183 & -404 & 111 \\
4& -26 & 524 & -2263 & 4089 & -3756 & 1833 & -401 \\
5& -3 & 283 & -2323 & 6321 & -7723 & 4378 & -933 \\
6& 0 & 90 & -1466 & 5416 & -7661 & 4488 & -867 \\
7& 0 & 15 & -576 & 2847 & -4455 & 2598 & -429 \\
8& 0 & 1 & -137 & 937 & -1585 & 902 & -118 \\
9& 0 & 0 & -18 & 188 & -339 & 186 & -17 \\
10& 0 & 0 & -1 & 21 & -40 & 21 & -1 \\
11& 0 & 0 & 0 & 1 & -2 & 1 & 0 \\\end{array}
\right),
\end{equation*}

\begin{equation*}
\hat{\bf N}^{\hat{Q}_3(5,-1)}_{[ 2]}=\left(
\begin{array}{c|ccccccc}
\hline
 k/m&6& 8 & 10 & 12 & 14 & 16 & 18 \\
\hline
0& 63 & -358 & 847 & -1068 & 757 & -286 & 45 \\
1& 179 & -1229 & 3336 & -4656 & 3555 & -1415 & 230 \\
2& 133 & -1190 & 3788 & -5779 & 4555 & -1773 & 266 \\
3& -16 & 535 & -2900 & 6863 & -8219 & 4870 & -1133 \\
4& 35 & 1163 & -9590 & 26003 & -32079 & 18484 & -4016 \\
5& 187 & -1525 & -205 & 13667 & -24632 & 15586 & -3078 \\
6& 180 & -4658 & 23587 & -47367 & 48092 & -26221 & 6387 \\
7& 74 & -4770 & 39338 & -110557 & 139417 & -81109 & 17607 \\
8& 14 & -2697 & 34551 & -120755 & 168332 & -99202 & 19757 \\
9& 1 & -918 & 19232 & -82433 & 123868 & -72781 & 13031 \\
10& 0 & -187 & 7117 & -37832 & 60532 & -35101 & 5471 \\
11& 0 & -21 & 1750 & -11889 & 20092 & -11412 & 1480 \\
12& 0 & -1 & 275 & -2528 & 4485 & -2481 & 250 \\
13& 0 & 0 & 25 & -348 & 645 & -346 & 24 \\
14& 0 & 0 & 1 & -28 & 54 & -28 & 1 \\
15& 0 & 0 & 0 & -1 & 2 & -1 & 0 \\\end{array}
\right),
\end{equation*}
\begin{equation*}
\hat{\bf N}^{\hat{Q}_3(5,-1)}_{[1,1]}=\left(
\begin{array}{c|ccccccc}
\hline
 k/m&6& 8 & 10 & 12 & 14 & 16 & 18 \\
\hline
 0&-39 & 224 & -539 & 696 & -509 & 200 & -33 \\
 1&-60 & 515 & -1584 & 2400 & -1942 & 805 & -134 \\
 2&58 & -36 & -590 & 1402 & -1218 & 422 & -38 \\
 3&119 & -961 & 3294 & -6354 & 6967 & -3957 & 892 \\
 4&-3 & -480 & 4118 & -12007 & 15574 & -9181 & 1979 \\
 5&-83 & 1206 & -3352 & 1784 & 2342 & -2160 & 263 \\
 6&-51 & 2093 & -13247 & 31322 & -35700 & 20263 & -4680 \\
 7&-12 & 1544 & -15851 & 49425 & -65504 & 38489 & -8091 \\
 8&-1 & 635 & -10816 & 42171 & -61294 & 36170 & -6865 \\
 9&0 & 150 & -4684 & 22866 & -35747 & 20882 & -3467 \\
 10&0 & 19 & -1311 & 8223 & -13675 & 7830 & -1086 \\
 11&0 & 1 & -230 & 1958 & -3437 & 1915 & -207 \\
 12&0 & 0 & -23 & 297 & -547 & 295 & -22 \\
 13&0 & 0 & -1 & 26 & -50 & 26 & -1 \\
 14&0 & 0 & 0 & 1 & -2 & 1 & 0 \\
\end{array}\right).
\end{equation*}

\bibliographystyle{elsarticle-num}
\bibliography{HWF}

\end{document}